\documentclass[11pt,a4paper,twocolumn]{article}

\usepackage[utf8]{inputenc}
\usepackage[T1]{fontenc}
\usepackage{lmodern}
\usepackage{microtype}
\usepackage[margin=2.0cm,top=2.4cm,bottom=2.4cm]{geometry}
\usepackage{graphicx}
\usepackage{xcolor}
\usepackage{booktabs}
\usepackage{tabularx}
\usepackage{amsmath,amssymb}
\usepackage{natbib}
\usepackage{hyperref}
\usepackage{cleveref}
\usepackage{caption}
\usepackage{subcaption}
\usepackage{enumitem}
\usepackage{xurl}
\usepackage{float}
\usepackage{multirow}
\usepackage{placeins}
\usepackage{tikz}
\usetikzlibrary{positioning,arrows.meta,shapes.geometric,calc,backgrounds,fit,shapes.misc}

\newcommand{\doi}[1]{\href{https://doi.org/#1}{\texttt{doi:#1}}}

\hypersetup{
  colorlinks=true,
  linkcolor={blue!60!black},
  citecolor={green!50!black},
  urlcolor={blue!70!black},
}

\title{%
  \textbf{CMIP-Forge: An Agentic System that Retrieves,\\
  Computes, and Self-Reviews Climate Science}%
}

\author{%
  Dmitrii Pantiukhin\textsuperscript{1,*}, Boris Shapkin\textsuperscript{1}, Ivan Kuznetsov\textsuperscript{1},\\
  Thomas Jung\textsuperscript{1}, and Nikolay Koldunov\textsuperscript{1}\\[4pt]
  \small\textsuperscript{1}Alfred Wegener Institute, Helmholtz Centre for Polar and Marine Research, Bremerhaven, Germany\\
  \small\textsuperscript{*}Corresponding author: \texttt{dmitrii.pantiukhin@awi.de}
}

\date{}

\begin{document}
\maketitle

\begin{abstract}
The Coupled Model Intercomparison Project Phase~6 (CMIP6) has generated thousands of peer-reviewed publications documenting model configurations, evaluation procedures, emergent constraints, and projection uncertainties.
As the community transitions toward CMIP7, efficiently extracting and operationalizing this unstructured knowledge alongside live data analysis represents a critical bottleneck.
Here we present \textbf{CMIP-Forge}, a hybrid retrieval-augmented generation (RAG) and autonomous analysis system that bridges the gap between scientific literature and Earth System Grid Federation (ESGF) data archives.
The system pairs a curated corpus of 6{,}581 CMIP6-related open-access publications (101{,}828 indexed chunks) with an agentic analytical pipeline in which a tool-augmented worker plans and executes cloud-native Python workflows over live climate data, while a configurable panel of independent reviewer models audits its methodology end to end.
Crucially, CMIP-Forge introduces a multi-layered \textit{Defense-in-Depth} architecture designed to enforce physical and methodological invariants. Rather than relying solely on prompt engineering, the system implements executable enforcement through Abstract Syntax Tree (AST) static analysis, audited scientific primitives, and an autonomous \textit{adversarial peer-review protocol}.
We demonstrate the system's capabilities through a series of end-to-end autonomous research pipelines spanning atmospheric teleconnections, ocean dynamics, regional extremes, and global warming projections.
The results show that an agentic analysis system grounded in peer-reviewed literature, constrained by automated code guardrails, and audited by an independent adversarial review loop can complete complex climate-research workflows autonomously. The same experiments expose concrete failure modes of the review loop (sycophantic regression, REVISE verdicts that are never resolved, and the submission of stub code for review), each diagnosable from the immutable telemetry and provenance record released with the article.

\medskip
\noindent\textbf{Keywords:} CMIP6, retrieval-augmented generation, large language models, autonomous agents, adversarial peer review, climate modeling, bias correction
\end{abstract}

\section{Introduction}\label{sec:intro}

The Coupled Model Intercomparison Project Phase~6 (CMIP6) has been the largest coordinated climate modeling effort to date, engaging more than 30 modeling groups worldwide and generating output across over 20 CMIP6-Endorsed Model Intercomparison Projects (MIPs) spanning scenarios from deep paleoclimate to idealized future forcing pathways \citep{Eyring2016}.
The scientific output of this effort is vast: a body of papers spanning more than a decade records how individual models were configured and tuned, how they were evaluated against observations, and how the resulting projection spreads should be interpreted.
Yet as the climate science community transitions toward CMIP7, a fundamental scalability challenge has emerged: the cumulative body of knowledge from CMIP6, spanning model descriptions, variable definitions, experiment protocols, and inter-model comparisons, presents a substantial barrier for any individual researcher or working group attempting systematic review.

This knowledge reuse problem is acute.
CMIP7 does not begin from zero; it inherits the infrastructure, parameterization experience, and documented biases of its predecessor.
Climate models that participated in CMIP6 carry forward tuning histories, grid configurations, and known systematic errors that are extensively documented in the CMIP6 literature.
Researchers designing new experiments, calibrating updated model versions, or interpreting projection spreads must locate and synthesize this prior art, a task that has traditionally relied on manual literature review and institutional memory.

The volume of relevant scientific output compounds this challenge.
Conservative estimates place the number of open-access CMIP6-related publications alone at over 7{,}000 papers spanning more than a decade of activity; the total corpus, including paywalled journals, is substantially larger.
These publications are distributed across hundreds of journals with varying access models, use inconsistent terminology for equivalent physical quantities, and frequently reference model configurations that have undergone multiple revision cycles.

This fragmentation of knowledge creates a critical bottleneck not just for model developers, but particularly for downstream users of climate data.
While the core physical science modeling community (such as the IPCC Working Group~I) maintains deep familiarity with the strengths, weaknesses, and structural biases of individual CMIP6 models, the sheer volume and fragmentation of this knowledge makes cross-community transfer inherently difficult.
Researchers in impacts, adaptation, and vulnerability (e.g., IPCC~WG~II) or mitigation strategy (WG~III) must often navigate complex model ensembles with limited access to the specialized documentation required to select appropriate models or apply necessary bias corrections for their specific application domains.

The emergence of Large Language Models (LLMs) as scientific reasoning engines has opened a fundamentally new approach to this knowledge synthesis problem.
LLMs have evolved from text generators into autonomous agents that reason over problems, decompose tasks, and invoke external tools \citep{Boiko2023,Schick2023}.
This shift has given rise to Multi-Agent Systems (MAS) in which complex problems are partitioned across specialized agents \citep{Hong2024,Guo2024}.
Within geosciences and climate science, this progression has driven rapid innovation.
Domain-specific foundation models such as K2 \citep{Deng2024} and ClimateGPT \citep{Thulke2024} have established stable baselines for Earth science knowledge extraction, while frameworks like GeoAgent \citep{Chen2024} have demonstrated the capability of LLMs to conduct autonomous geospatial data analysis through code interpretation.
Concurrently, specialized agentic systems have moved into operational deployment: ClimSight \citep{Koldunov2024,Kuznetsov2025} demonstrated that augmenting LLMs with localized climate model output and RAG can deliver actionable, location-specific climate assessments, while PANGAEA-GPT \citep{Pantiukhin2025a,Pantiukhin2025b} established a hierarchical multi-agent framework for autonomous data discovery and analysis in geoscientific data archives.

However, existing agent systems in climate science operate primarily on structured data resources (model output fields, reanalysis grids, and curated dataset repositories).
None provides systematic access to the unstructured scientific literature that documents \emph{how} these data were produced, what their known limitations are, and how they should be interpreted.
What is missing is a system that merges literature-grounded domain knowledge with live data analysis capabilities.
Moreover, single-agent code generation in the Earth sciences is notoriously prone to subtle physical errors (e.g., misinterpreting accumulated variables as rate fluxes, or mishandling non-Gregorian calendars), which can silently corrupt climate projections.

To close this gap, we introduce \textbf{CMIP-Forge}, an engine that couples retrieval over the CMIP6 literature with autonomous, tool-driven data analysis.
The system integrates a curated corpus of over 6{,}500 CMIP6-related scientific publications with a tool-augmented LLM agent capable of searching datasets, executing Python-based analyses, and retrieving ERA5 reanalysis data directly from the Copernicus Climate Data Store (CDS).
To strengthen scientific rigor, it introduces an \emph{autonomous adversarial peer-review mechanism}: analytical pipelines proposed by the primary worker agent are automatically subjected to physics-based critiques by independent reviewer agents.
By combining literature-grounded domain knowledge with integrated observational reanalysis access and iterative computational capabilities, the system enables a workflow in which a researcher can submit a natural language prompt and receive a detailed, literature-informed climate assessment.

\section{Methods}\label{sec:methods}

\subsection{System Overview}\label{sec:overview}

\begin{figure*}[!t]
  \centering
  \begin{tikzpicture}[
    font=\sffamily\footnotesize,
    >={Stealth[length=1.6mm,width=1.4mm]},
    every node/.style={align=center,inner sep=2pt},
    box/.style={rectangle,draw=black,fill=white,line width=0.22mm,
      text width=2.4cm,minimum height=2cm,inner sep=2mm,font=\sffamily\scriptsize},
    wide/.style={rectangle,draw=black,fill=white,line width=0.22mm,
      text width=4cm,minimum height=1.3cm,inner sep=2mm,font=\sffamily\scriptsize},
    header/.style={rectangle,draw=black,fill=white,line width=0.4mm,
      text width=13.8cm,minimum height=1.3cm,inner ysep=3mm,inner xsep=4mm,
      font=\sffamily\small\bfseries},
    input/.style={rectangle,draw=black,fill=white,line width=0.22mm,
      text width=4.5cm,minimum height=0.85cm,font=\sffamily\small\bfseries},
    final/.style={rectangle,draw=black,fill=white,line width=0.4mm,
      text width=4.5cm,minimum height=0.85cm,font=\sffamily\small\bfseries},
    arr/.style={->,line width=0.22mm,black},
    fb/.style={->,line width=0.22mm,dashed,black!60},
  ]

  \node[input] (user) at (-0.8,0) {User \\[0.4mm] \scriptsize\normalfont natural-language prompt};

  \node[header,below=4mm of user] (agent) {%
    Worker Agent \;\textbar\; LangGraph ReAct \;\textbar\; Foundation Model\\[1mm]
    \scriptsize\normalfont 14 tools \quad 9 invariants \quad 7 failure-mode exemplars \quad Empirical Defiance Protocol%
  };

  \node[box,below=6mm of agent.south,xshift=-5.3cm] (lit) {%
    \textbf{Literature}\\[1.5mm]
    literature\_search\\[0.5mm] citation\_graph\\[0.5mm] methodology\_check};
  \node[box,right=3mm of lit] (data) {%
    \textbf{Data access}\\[1.5mm]
    datasets\_search\\[0.5mm] datasets\_access\\[0.5mm] cmip6\_adviser\\[0.5mm] era5\_monthly};
  \node[box,right=3mm of data] (comp) {%
    \textbf{Compute}\\[1.5mm]
    python\_repl\\[0.5mm] analysis\_guide};
  \node[box,right=3mm of comp] (mem) {%
    \textbf{Memory}\\[1.5mm]
    save\_to\_memory\\[0.5mm] forget};
  \node[box,right=3mm of mem] (qa) {%
    \textbf{Quality}\\[1.5mm]
    review\_figure\\[0.5mm] reviewer\_1\\[0.5mm] reviewer\_2};

  \node[box,below=6mm of lit] (qd) {%
    \textbf{Qdrant index}\\[1.5mm]
    4 collections\\[0.5mm] 103{,}596 points\\[0.5mm] dense + BM25};
  \node[box,below=6mm of data] (esgf) {%
    \textbf{ESGF / Pangeo}\\[1.5mm]
    1{,}313 variables\\[0.5mm] 132 sources\\[0.5mm] 323 experiments};
  \node[box,below=6mm of comp,minimum height=1.3cm] (lint) {%
    \textbf{AST Linter}\\[1mm]
    \scriptsize 4 rules; critical block};
  \node[box,below=6mm of qa,minimum height=1.3cm] (rev) {%
    \textbf{Reviewer Panel}\\[1mm]
    \scriptsize Gemini, Claude, GPT};

  \node[wide,below=7mm of lint,xshift=1cm,text width=3.5cm] (repl) {%
    \textbf{Sandboxed Python REPL}\\[1mm]
    audited primitives \quad runtime telemetry};

  \node[final,below=5mm of repl] (out) {%
    Figures \ + \ telemetry JSON};

  \draw[arr] (user) -- (agent);
  \draw[arr] (agent.south -| lit.north) -- (lit.north);
  \draw[arr] (agent.south -| data.north) -- (data.north);
  \draw[arr] (agent.south -| comp.north) -- (comp.north);
  \draw[arr] (agent.south -| mem.north) -- (mem.north);
  \draw[arr] (agent.south -| qa.north) -- (qa.north);

  \draw[arr] (lit) -- (qd);
  \draw[arr] (data) -- (esgf);
  \draw[arr] (comp) -- (lint);
  \draw[arr] (qa) -- (rev);

  \draw[arr] (lint.south) -| (repl.north);
  \draw[arr] (repl) -- (out);

  \draw[fb] (rev.north east) |- node[pos=0.55,right,font=\sffamily\tiny,black!70,xshift=0.5mm]{verdict} (agent.east);
  \draw[fb] (repl.east) -| node[pos=0.45,above,font=\sffamily\tiny,black!70,yshift=0.5mm]{telemetry} (rev.south);

  \end{tikzpicture}
  \caption{\textbf{CMIP-Forge agentic architecture.} A user prompt is consumed by a ReAct worker agent (LangGraph) whose system prompt encodes nine geophysical invariants, seven failure-mode exemplars, and the Empirical Defiance Protocol. The agent has access to fourteen tools grouped into five categories. Literature retrieval is backed by a Qdrant hybrid-search index (dense Gemini Embedding 2 plus sparse BM25 across 101{,}828 literature chunks in the primary collection, alongside three CMIP6 facet collections); data access resolves natural-language facet queries against the ESGF / Pangeo zarr catalog and the Copernicus ERA5 CDS. Every code block emitted by the agent passes through an AST static-analysis linter; critical violations block execution. Code that passes runs in a sandboxed Python REPL pre-loaded with audited scientific primitives, with runtime telemetry automatically captured. The complete payload (code, figures, telemetry) is submitted to a panel of adversarial reviewer agents drawn from independent foundation-model families; reviewer verdicts (ACCEPT, REVISE, REJECT) flow back to the worker, which iterates until the panel emits a passing verdict. All session traces are released verbatim as immutable provenance.}
  \label{fig:system_ui}
\end{figure*}
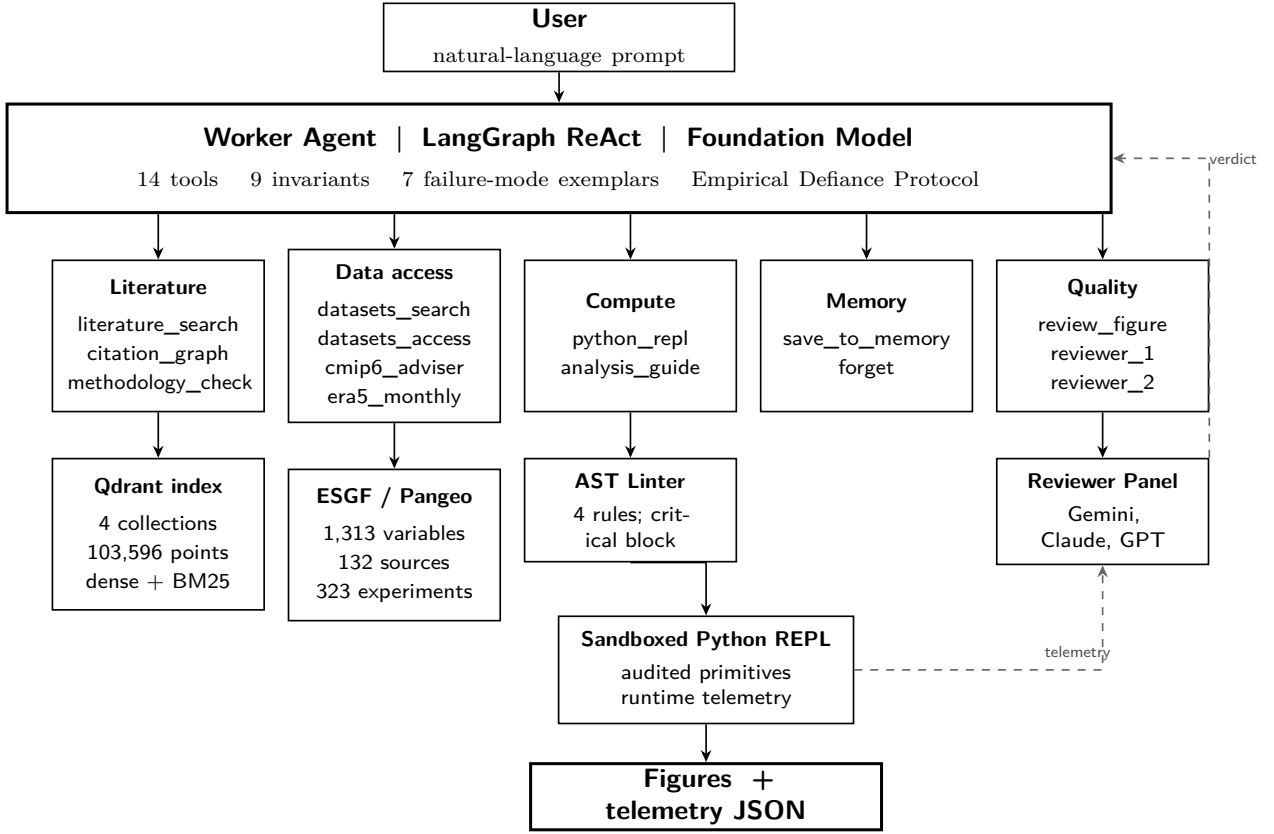

CMIP-Forge integrates scientific literature retrieval, live climate data access, and autonomous code execution into a single interactive system, enabling end-to-end climate analyses (from hypothesis formulation through publication-quality visualization) within a unified agent-driven workflow. These capabilities are wrapped in a code-guardrail and adversarial-review pipeline; \cref{fig:system_ui} shows the full architecture. Its core data-and-reasoning subsystems are:
\begin{enumerate}[nosep]
  \item A \textbf{literature search engine} (Retrieval-Augmented Generation) over 6{,}581 CMIP6 papers, so the agent can look up methods, known biases, and physical constraints before writing any code.
  \item A \textbf{CMIP6 metadata resolver} that translates natural language (e.g., ``sea surface temperature from MPI'') into the exact ESGF facets needed to locate datasets.
  \item An \textbf{ERA5 reanalysis module} that downloads observational fields directly from the Copernicus Climate Data Store, enabling model--observation comparison within a single session.
  \item A \textbf{ReAct reasoning agent} \citep{Yao2023} that plans multi-step analyses, calls the above tools, executes Python code in a sandboxed environment, and iterates until the answer is complete.
\end{enumerate}
The system is deployed as a web application (React frontend, FastAPI backend) with real-time streaming via Server-Sent Events.

A key lesson from early deployments was that \emph{LLMs can generate syntactically correct code that is physically meaningless}: for example, applying image-processing filters to geophysical grids or computing spatial means without area weighting. Prompt instructions alone proved insufficient to prevent such errors. We therefore introduced a four-layer \textit{Defense-in-Depth} architecture that shifts the burden of correctness from the language model's reasoning to executable programmatic enforcement:
\begin{enumerate}[nosep]
  \item \textbf{Static code analysis.} An Abstract Syntax Tree (AST) linter inspects every code block \emph{before} execution. It enforces four categories of geophysical guardrails: (i)~\emph{data-fabrication detection}, which raises a critical-severity flag when a CMIP6-named variable (e.g.\ \texttt{awi\_tos}, \texttt{mpi\_sst}) is populated from \texttt{np.random.*} or similar synthetic sources (i.e.\ the agent silently inventing model output when a real fetch fails); (ii)~\emph{image-processing operators on geophysical grids}, where \texttt{sobel}/\texttt{Laplacian}/\texttt{Scharr} kernels raise a blocking exception because they compute index-space rather than physical-distance gradients; (iii)~\emph{unweighted spatial means} over lat/lon dimensions, which warn the agent to use cosine-latitude weighting; and (iv)~\emph{additive in-place scalar shifts} (\texttt{data\,-=\,bias}) and explicit axis-limit overrides, both of which often mask guardrail-hacking attempts to force baselines or hide data outside the chosen range. Full rule list, severity policy, and integration with the worker--reviewer loop are in Section~S0 of the Supplementary Materials.
  \item \textbf{Audited scientific primitives.} The analysis sandbox is pre-loaded with verified helper functions (area-weighted means with mask alignment, Earth-radius-aware gradients, continuous DJF seasonal aggregation) so the agent uses tested building blocks instead of improvising its own.
  \item \textbf{Runtime telemetry.} Every code execution automatically prints summary statistics (minimum, mean, maximum, array dimensions) to the standard output stream. These numbers are injected verbatim into the reviewer context, providing an empirical anchor that is harder to hallucinate away.
  \item \textbf{Empirical Defiance Protocol.} The worker agent is explicitly instructed to reject reviewer critiques that contradict the observed telemetry, even when the critique sounds authoritative, countering the well-documented tendency of LLMs to defer to confident-sounding but incorrect feedback.
\end{enumerate}

\subsection{Literature Corpus Construction}\label{sec:corpus}

\subsubsection{Corpus Scope and Acquisition}

The literature corpus comprises 6{,}581 unique scientific publications related to CMIP6 climate science; the complete list, with DOIs and licence metadata, is distributed alongside the code and data release.
Papers were identified through systematic queries of the OpenAlex academic metadata API, targeting publications that reference CMIP6 models, experiments, variables, or scenarios.
Of the 7{,}162 publications identified, 7{,}106 (99.2\% of identified) were successfully downloaded from publisher endpoints and open-access repositories (PubMed Central, Semantic Scholar, CORE); the remaining 56 were blocked by publisher access restrictions.
After parsing and quality filtering, 6{,}581 documents (92.6\%) yielded usable text and were indexed into the RAG corpus. The 525 excluded documents contained corrupted PDFs, image-only scans, or produced no qualifying text chunks after boilerplate removal.

The indexed publications were retrieved from open-access endpoints, with 99.4\% confirmed as open access against the OpenAlex API. The RAG system stores only short text chunks (up to 1{,}000 tokens) and their vector embeddings, not full documents.

\subsubsection{Document Parsing}

Raw PDF documents were processed using the MinerU document understanding framework \citep{Wang2024mineru,Dong2026minerudiffusion,Wang2026mineru2}, deployed on dual NVIDIA A100 GPUs with a vLLM inference backend.
MinerU employs a compact dedicated vision-language model that performs layout-aware parsing of each page, distinguishing headings, body paragraphs, tables, mathematical formulas, and figure captions while preserving their hierarchical section paths.
Unlike flat text extraction tools, which discard structural information, MinerU reconstructs the logical document tree, including multi-column layouts, inline equations, and nested subsections, enabling section-aware processing in downstream stages.
The raw output passed through a preprocessing stage that stripped publisher-specific boilerplate, linked figure images to their captions, and corrected systematic OCR artifacts.

\subsubsection{Section-Aware Chunking}

Parsed documents are segmented into semantically coherent chunks by a custom section-aware chunker (V6) that operates on the MinerU JSON output.
Key design decisions include a target chunk size of up to 1{,}000 tokens with a hard cap of 1{,}200 tokens for table chunks, a minimum quality threshold of 30 tokens (chunks below 80 tokens are merged with adjacent content), 5\% overlap ratio between consecutive chunks, abstract isolation as standalone chunks, and automatic filtering of academic boilerplate sections (references, acknowledgements, etc.) via a curated exclusion list of $>$80 normalized section headings.

The chunking pipeline was systematically refined to ensure high data quality: removal of HTML/URN boilerplate, suppression of OCR artifacts, content-hash deduplication, and context-dependent OCR corrections.
The pipeline produced 101{,}828 chunks across the 6{,}581 papers, with a mean of 15.5 chunks per paper.

\subsubsection{Embedding}

Each text chunk is embedded into a 768-dimensional dense vector using Google's Gemini Embedding~2 model (\texttt{gemini-embedding-2-preview}), with asymmetric encoding applied separately for documents during indexing and for queries at search time.
In addition, a sparse BM25 (Best Matching~25, a term-frequency ranking function) vector is computed for each chunk via the FastEmbed library~\citep{FastEmbed2024}.
We use both representations because semantic embeddings alone tend to conflate CMIP6-specific identifiers (model source IDs, variable short names, experiment labels) that differ by only a few characters but carry distinct physical meaning, whereas BM25 retrieves them exactly.
Both vector types are stored in Qdrant \citep{Qdrant2024}, an open-source vector database with native support for hybrid dense--sparse search.

\subsection{Vector Database and Hybrid Search}\label{sec:search}

\subsubsection{Qdrant Index Architecture}

The embedded corpus is indexed in a Qdrant vector database configured for hybrid retrieval.
The primary literature collection (101{,}828 indexed chunks) stores both dense (768-dimensional cosine) and sparse (BM25) vector representations for each chunk, alongside structured metadata payloads containing: paper DOI, title, publication year, journal name, section path, chunk type, and quality tier.

\subsubsection{Hybrid Search Pipeline}

Literature retrieval is driven entirely by the LLM agent, which decomposes the user's request into one or more focused search queries and invokes the search tool multiple times as needed; the raw user input never reaches the search engine directly.
Each agent-generated query then passes through a three-stage pipeline:

\begin{enumerate}[nosep]
  \item \textbf{Dual-channel query encoding}: The query is simultaneously embedded as a dense vector (Gemini Embedding) and a sparse BM25 vector.
  \item \textbf{Reciprocal Rank Fusion (RRF)}: Qdrant executes parallel prefetch queries against the dense and sparse indices (50 candidates per channel), then fuses the ranked lists using Reciprocal Rank Fusion (RRF) \citep{Cormack2009}.
  \item \textbf{Neural reranking}: The fused candidate set is reranked using Google's Vertex AI Ranking API (\texttt{semantic-ranker-default@latest}); if the API call fails, the pipeline gracefully degrades to the RRF-fused top-$k$ without an additional reranking step.
\end{enumerate}

The pipeline supports structured filtering by publication year, journal, quality tier, chunk type, and DOI exclusion lists, all specified by the agent as tool parameters. To ensure correct cosine similarity matching, the RAG threshold is explicitly calibrated to 0.5.

\subsubsection{Citation Graph}

Keyword-based search alone cannot capture the lineage of scientific ideas, for example tracing how a bias documented in an early CMIP6 evaluation was later addressed in a follow-up study. To support this, a citation graph is constructed from the corpus metadata, representing inter-paper citation relationships as a directed graph. The agent can traverse this graph in both directions: forward (which later papers cited a given study?) and backward (what prior work did it build on?), with results ranked by global citation count. This enables the agent to follow methodological chains across the literature rather than relying on isolated keyword matches.

\subsection{CMIP6 Dataset Discovery}\label{sec:facets}

Alongside literature search, the agent can locate and retrieve specific CMIP6 datasets from the Earth System Grid Federation (ESGF) archive.
ESGF organises data along a set of controlled-vocabulary facets (variable, source model, experiment, frequency, realm, and others) that must be specified exactly to construct a valid query.
However, while smaller facets such as frequency, realm, and resolution contain only a handful of values and can be passed directly as enumerated literals in the tool schema, the three largest facets (1{,}313 distinct variables, 323 experiments, and 132 model source IDs) are far too large to enumerate within an LLM context window at every request, both for cost and latency reasons.

We address this by placing the three largest facets (variables, experiments, and sources) on a dedicated vector retrieval layer, stored as three separate Qdrant collections.
A key challenge was that the native ESGF metadata descriptions are extremely terse and inconsistent: many variables lack meaningful comments, experiment descriptions are written as single-clause protocol references, and model source entries consist of bare component lists with no scientific context.
To close this semantic gap, we used Gemini~3.1~Pro to rewrite every variable, experiment, and source metadata entry into a 200--260 word search-optimized paragraph.
Each rewriting prompt included structured examples, mandatory disambiguation from confusable entries, cross-references to related quantities, and a quality-control loop that rejected outputs containing filler phrases, anaphoric pronouns, or low vocabulary diversity.
The rewritten descriptions were then embedded with the same Gemini Embedding~2 model used for the literature corpus.

When the agent invokes the dataset search tool, the following pipeline executes:
\begin{enumerate}[nosep]
  \item The agent passes natural-language sub-queries for each facet (e.g., \texttt{variable\_query="sea surface temperature"}, \texttt{source\_query="MPI"}).
  \item Each sub-query is embedded and matched against its corresponding Qdrant collection via cosine similarity, with exact-match injection and query-term boosting applied during re-ranking.
  \item The top-$k$ candidates from all facets are assembled into a dynamic schema, and a single LLM call selects the best-matching CMIP6 identifiers from the ranked options.
  \item The resolved facet values are used to query the ESGF Search API, returning dataset counts, available models, and download links.
\end{enumerate}
For batch requests (e.g., searching for multiple variables across the same model), vector searches run independently per query while the facet selection is batched into a single LLM call to minimise latency. If the ESGF pipeline returns an error, the system enforces a hard-fail protocol: the agent receives an unambiguous failure signal and is blocked from proceeding with fabricated dataset metadata.

\subsection{Agent Architecture}\label{sec:agent}

\subsubsection{ReAct Agent}

The CMIP-Forge agent is implemented as a LangGraph ReAct agent, a compiled directed cyclic graph implementing the ReAct (Reasoning + Acting) paradigm \citep{Yao2023}.
The agent receives a natural language query, reasons about the required information, selects and invokes tools, observes results, and iterates until it can formulate a grounded response. The worker agent's system prompt is strictly structured to mitigate sycophancy and logic errors. It encodes mandatory geospatial and mathematical invariants, requires summary statistics to be printed before plotting, provides few-shot exemplars of known failure modes (such as heatwave semantics, gradient physics, and free-running variability), and applies the Empirical Defiance Protocol.

\subsubsection{Tool Registry}

The agent has access to fourteen specialized tools (\cref{tab:tools}).

\begin{table*}[t]
  \centering
  \caption{CMIP-Forge tool registry. Tools are grouped by function: literature retrieval, CMIP6 data access, computation, and quality assurance. Names are shortened for display; registered identifiers carry prefixes (e.g.\ \texttt{cmip6\_literature\_search}, \texttt{retrieve\_era5\_monthly}, \texttt{get\_analysis\_guide}).}
  \label{tab:tools}
  \small
  \begin{tabularx}{\textwidth}{@{}l l X@{}}
    \toprule
    \textbf{Category} & \textbf{Tool} & \textbf{Function} \\
    \midrule
    \multirow{3}{*}{\emph{Literature}}
      & \texttt{literature\_search}  & Hybrid dense--sparse retrieval over the 101K-chunk corpus; supports year, journal, and DOI filters with Vertex AI reranking. \\
      & \texttt{citation\_graph}     & Forward/backward citation traversal within the corpus, ranked by global citation count. \\
      & \texttt{methodology\_check}  & Methodology-focused RAG over the same corpus; invoked \emph{before} analysis code to verify procedures and known pitfalls. \\
    \midrule
    \multirow{4}{*}{\emph{Data access}}
      & \texttt{datasets\_search}    & Resolves natural-language queries to CMIP6 facets via metadata RAG (\S\ref{sec:facets}), then checks ESGF availability. \\
      & \texttt{datasets\_access}    & Direct ESGF availability check given pre-resolved facet values. \\
      & \texttt{cmip6\_adviser}      & Answers metadata questions about variables, models, and experiments via the enriched facet collections. \\
      & \texttt{era5\_monthly}       & Downloads ERA5 monthly reanalysis from the Copernicus CDS API; returns a NetCDF path for REPL analysis. \\
    \midrule
    \multirow{2}{*}{\emph{Compute}}
      & \texttt{python\_repl}        & Persistent sandboxed Python environment with session-isolated workspace and automatic figure capture. \\
      & \texttt{analysis\_guide}     & Deterministic methodological templates (workflows, quality checklists, pitfalls) for 20+ CMIP6 analysis patterns. \\
    \midrule
    \multirow{2}{*}{\emph{Memory}}
      & \texttt{save\_to\_memory}    & Writes a small key-value entry to a session-scoped blackboard for cross-tool persistence. \\
      & \texttt{forget}              & Removes a blackboard entry by key when it becomes stale or capacity is reached. \\
    \midrule
    \multirow{3}{*}{\emph{QA}}
      & \texttt{review\_figure}      & Multimodal visual inspection of figures in three modes: \emph{correct}, \emph{describe}, and \emph{qa} (pass/fail). \\
      & \texttt{reviewer\_1}         & Adversarial peer review of the full analysis by an LLM drawn from a configurable pool (Gemini, Claude, GPT). \\
      & \texttt{reviewer\_2}         & Second independent reviewer, configured per request to use a structurally different model. \\
    \bottomrule
  \end{tabularx}
\end{table*}

\subsubsection{Session Management and Executable Guardrails}

Each user session runs in an isolated Python REPL, identified by a UUID, with its own filesystem sandbox for generated artifacts. A per-session lock serializes concurrent requests to prevent race conditions, and REPL state persists across tool calls so multi-step analyses do not reload data.

An AST linter statically analyzes each code block before execution. It applies hard gates, for example raising an exception when image-processing gradient kernels (Sobel, Laplacian) are applied to geophysical grids, as well as soft warnings, such as on unweighted spatial means or scalar bias shifts. The REPL namespace is also pre-loaded with audited primitives covering the most common silent-error sources: mask-aligned area-weighted averaging, Earth-radius-scaled (physical-space) gradients on latitude--longitude grids, cross-year-consistent boreal-winter (DJF) aggregation, and machine-readable figure-metadata extraction for visual quality audits.

\subsubsection{Methodology Grounding and Visual Quality Assurance}

Three additional tools support methodological integrity.
A dedicated methodology retrieval tool queries the same 101K-chunk literature corpus used for general search, but with a methodology-specific focus: the agent invokes it \emph{before} writing analysis code to retrieve peer-reviewed guidance on unit conversion procedures, statistical methods, physical constraints, and known pitfalls for specific variables or regions.
A metadata adviser tool answers factual questions about model component descriptions, experiment protocols, and variable definitions by querying the enriched facet collections described in \S\ref{sec:facets}, without requiring a full dataset search.
An analysis guide tool provides deterministic, pre-authored methodological templates covering over twenty standard CMIP6 workflows (data loading, spatial subsetting, anomaly computation, EOF analysis, multi-model comparison, and others), each structured as a workflow with quality checklists and documented pitfalls.

After the Python REPL produces a figure, a multimodal visual inspection tool passes the rendered image to a vision-capable LLM in one of three modes: \emph{correct} (identify visual and scientific issues), \emph{describe} (narrate the figure content for downstream reasoning), or \emph{qa} (binary pass/fail against specified criteria).
This provides a visual feedback loop that complements the code-level review performed by the adversarial reviewers.

\subsection{Autonomous Adversarial Peer Review}\label{sec:review}

A persistent challenge in LLM-driven autonomous data analysis is the silent propagation of methodological errors that compile without runtime exceptions but yield physically meaningless results.

To mitigate this, CMIP-Forge implements an asynchronous \textbf{Multi-Agent Peer Review} protocol governed by typed, machine-readable message contracts. Before a final analytical result is accepted, the worker agent submits a structured review payload containing the task description, methodology rationale, complete executable code, generated figures, and (critically) the runtime telemetry captured during execution (empirical statistics such as minimum, mean, maximum, and array dimensions).

Each reviewer slot is assigned, per request, a model from a configurable pool of frontier foundation models (Gemini~3.1~Pro, Claude Opus~4.x, GPT-5.5); in the use cases reported here the two reviewers were drawn from different model families (Claude Opus and GPT). We explicitly document that assigning the same model to both reviewers yields zero epistemic independence, as identical models inherit the same systematic biases and hallucination patterns, a direct consequence of the foundation-model homogenization risk identified by \citet{Bommasani2022}. Empirical work on multi-agent debate has confirmed this concern: \citet{Liang2024mad} identify \textit{Degeneration-of-Thought}, whereby an agent that is confident in its answer cannot generate novel corrective thoughts through self-reflection, motivating debate among epistemically independent agents.

Reviewers return structured critique reports in which each identified issue specifies a severity level, the contested claim, the type of supporting evidence, and a proposed fix. The reviewer system prompt strictly mandates trusting the empirical execution output over metadata assumptions. To counter sycophancy, the worker evaluates each critique against the observed runtime telemetry using the Empirical Defiance Protocol, rejecting mathematically destructive fixes when the execution statistics demonstrate physical validity. If valid issues are flagged, the worker engages in an autonomous self-correction loop until the reviewer panel emits a passing verdict.

\subsection{Frontend and Deployment}\label{sec:deploy}

The system is deployed as a web application:
the backend is a FastAPI server hosting the LangGraph agent with SSE streaming;
the frontend is a Vite/React single-page application providing a conversational interface with rendered Markdown, inline figure display, and RAG source attribution panels showing retrieved paper titles, DOIs, and relevance scores.

\section{Results}\label{sec:results}

\subsection{RAG Corpus Statistics}\label{sec:corpus_stats}

The constructed literature corpus provides extensive coverage of the CMIP6 scientific knowledge base (\cref{tab:corpus}).

\begin{table}[t]
  \centering
  \caption{CMIP-Forge corpus and index statistics.}
  \label{tab:corpus}
  \small
  \begin{tabular}{lr}
    \toprule
    \textbf{Metric} & \textbf{Value} \\
    \midrule
    Total papers parsed & 6{,}581 \\
    Total chunks indexed & 101{,}828 \\
    Mean chunks per paper & 15.5 \\
    Qdrant collections & 4 \\
    Total indexed points & 103{,}596 \\
    Confirmed open access (of indexed) & 99.4\% \\
    Embedding dimensionality & 768 \\
    \bottomrule
  \end{tabular}
\end{table}

The corpus was assembled via a multi-stage open-access ingestion pipeline. An initial bulk download pass retrieved PDFs from OpenAlex metadata links, capturing approximately 93\% of the target list. Subsequent recovery stages queried alternative open-access repositories (PubMed Central, Semantic Scholar, and CORE) to maximize coverage. The final pipeline achieved a 99.2\% retrieval rate from an initial target of 7{,}162 candidate papers identified through OpenAlex bibliometric queries filtered by CMIP6-related keywords and citation networks.

The corpus spans publications from over 120 journals covering the period 2016--2026, with the highest representation from core CMIP6 venues: \textit{Geoscientific Model Development}, \textit{Journal of Advances in Modeling Earth Systems}, \textit{Earth System Dynamics}, \textit{Journal of Climate}, \textit{Nature Climate Change}, \textit{Climate Dynamics}, and \textit{Geophysical Research Letters}. The temporal distribution peaks sharply in 2020--2023, coinciding with the main CMIP6 analysis and evaluation phase. Papers were filtered to retain only those with substantive CMIP6 content (model evaluation, scenario projections, or variable-specific analyses), excluding peripheral mentions and editorials.

\subsection{Retrieval Quality}\label{sec:retrieval}

A controlled, ground-truth-based benchmark of retrieval quality (Precision@$k$, nDCG, dense-vs-sparse-vs-hybrid+rerank ablation, latency profiling) requires a domain-specific evaluation set with annotated relevance judgements that is not yet available for this corpus. The downstream utility of the retrieval pipeline is instead exercised empirically through the seven end-to-end use cases reported in Section~\ref{sec:usecases}, in which every numerical claim, methodological choice, and DOI citation is traceable to specific retrieved chunks recorded in the per-use-case telemetry (Supplementary Materials). A formal retrieval benchmark on a forthcoming annotated CMIP6 query/relevance corpus is deferred to a companion methodological paper.

\subsection{Use Cases}\label{sec:usecases}

To illustrate end-to-end behaviour of CMIP-Forge on non-trivial scientific questions, we present a sequence of autonomous use cases spanning ocean dynamics, atmospheric teleconnections, and regional extremes. Each case follows the same protocol: a single natural-language prompt initiates a multi-step session in which the worker agent retrieves literature from the RAG corpus, discovers and downloads the relevant CMIP6 model output via ESGF, performs the analysis in the sandboxed Python environment, and produces publication-grade figures together with a written synthesis. Extended methodology, full agent telemetry, per-model diagnostics, and supplementary figures for every use case are provided in the Supplementary Materials.

\subsubsection{AMOC Deceleration and the Limits of European Shielding}\label{sec:uc_amoc}

\begin{figure*}[!b]
  \centering
  \includegraphics[width=\textwidth]{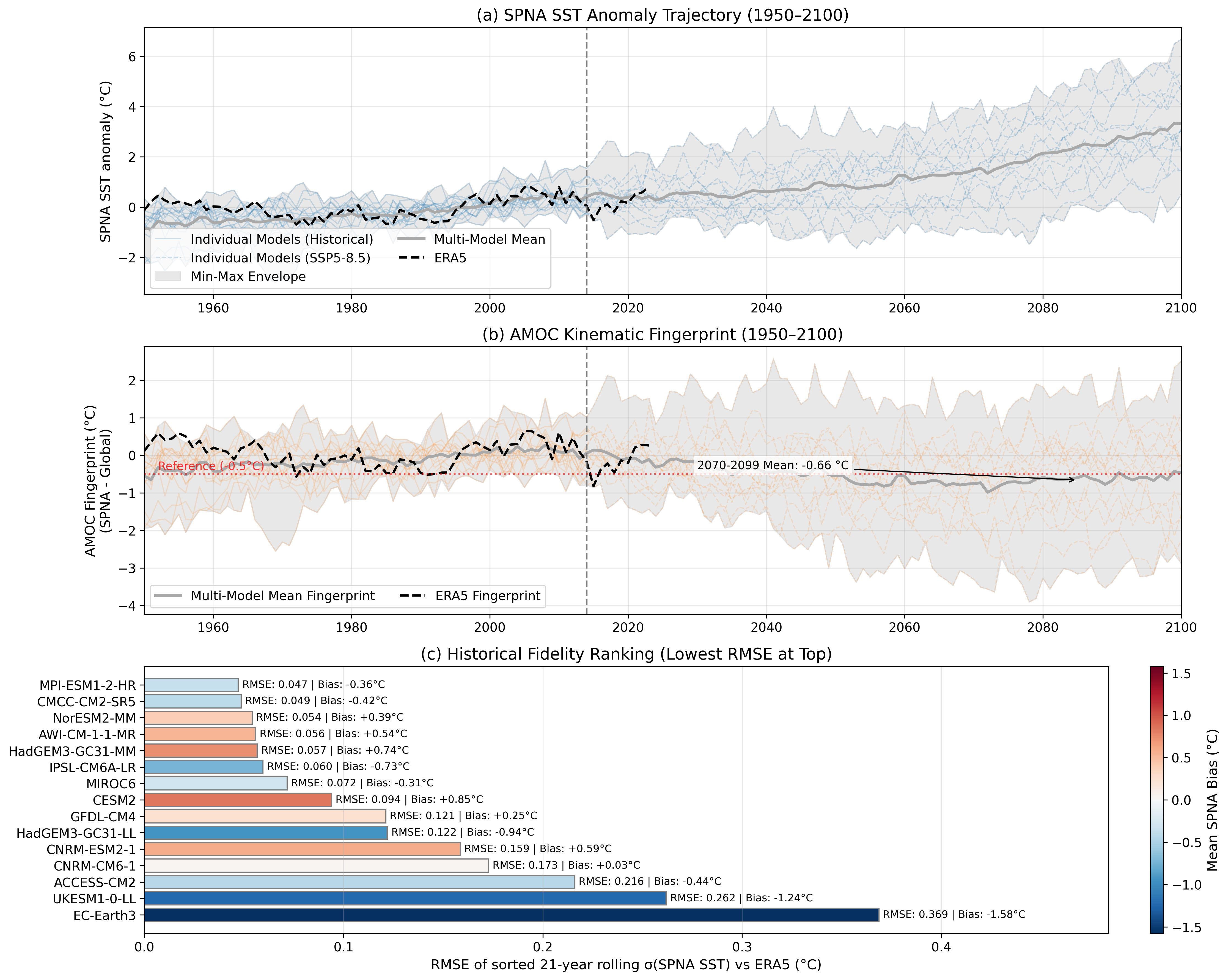}
  \caption{\textbf{Upstream oceanic diagnostic: AMOC kinematic fingerprint and 15-model historical fidelity ranking.} (a)~SPNA SST anomaly trajectories 1950--2100 under historical and SSP5-8.5 forcing, with the multi-model mean (MMM), min--max envelope, and ERA5 (1950--2023) as the observational reference. (b)~Kinematic fingerprint $\mathcal{F}(t)=\overline{\mathrm{SST}}_{\mathrm{SPNA}}-\overline{\mathrm{SST}}_{\mathrm{Global}}$; annotations mark the conventional $-0.5^\circ\mathrm{C}$ emergence threshold and the MMM 2070--2099 value of $-0.66^\circ\mathrm{C}$. (c)~Historical fidelity ranking of all 15 CMIP6 models against ERA5, sorted by the RMSE of the sorted 21-year rolling standard deviation of SPNA SST, with per-bar RMSE and signed mean SPNA bias annotations; bar colour encodes the bias on a diverging scale.}
  \label{fig:amoc_upstream}
\end{figure*}

The first use case targets the Atlantic Meridional Overturning Circulation (AMOC) and the long-standing hypothesis that its projected slowdown should partially buffer Northwest Europe against winter warming. From a single prompt, the agent assembled a paleoceanographic and dynamical literature synthesis, then operationalised the AMOC kinematic fingerprint as the difference between Subpolar North Atlantic (SPNA, 45\textdegree--65\textdegree N, 60\textdegree--10\textdegree W) and global-mean Sea Surface Temperature, following the standard advective-heat-deficit framing. Fifteen CMIP6 models were ranked against ERA5 (1950--2014) on the basis of their historical SPNA temperature variance and mean bias. The agent retained \texttt{MPI-ESM1-2-HR}, \texttt{AWI-CM-1-1-MR}, and \texttt{HadGEM3-GC31-MM} as the high-fidelity carry-forward ensemble and rejected \texttt{EC-Earth3} and \texttt{UKESM1-0-LL} as historical outliers carrying SPNA mean biases of $-1.58^\circ\mathrm{C}$ and $-1.24^\circ\mathrm{C}$ respectively (\cref{fig:amoc_upstream}, panel~c; full ranking and per-model statistics in Section~S1). Two ESGF member-availability substitutions were performed and logged autonomously across the chained sessions: \texttt{CESM2}~r1i1p1f1~$\rightarrow$~r11i1p1f1 in the SPNA-evaluation step, and \texttt{HadGEM3-GC31-MM}~$\rightarrow$~\texttt{-LL} in the downstream atmospheric-response step, the latter triggered by \texttt{HadGEM3-GC31-MM} not publishing \texttt{tas}/\texttt{ssp245} on ESGF.

\begin{figure*}[!b]
  \centering
  \includegraphics[width=\textwidth]{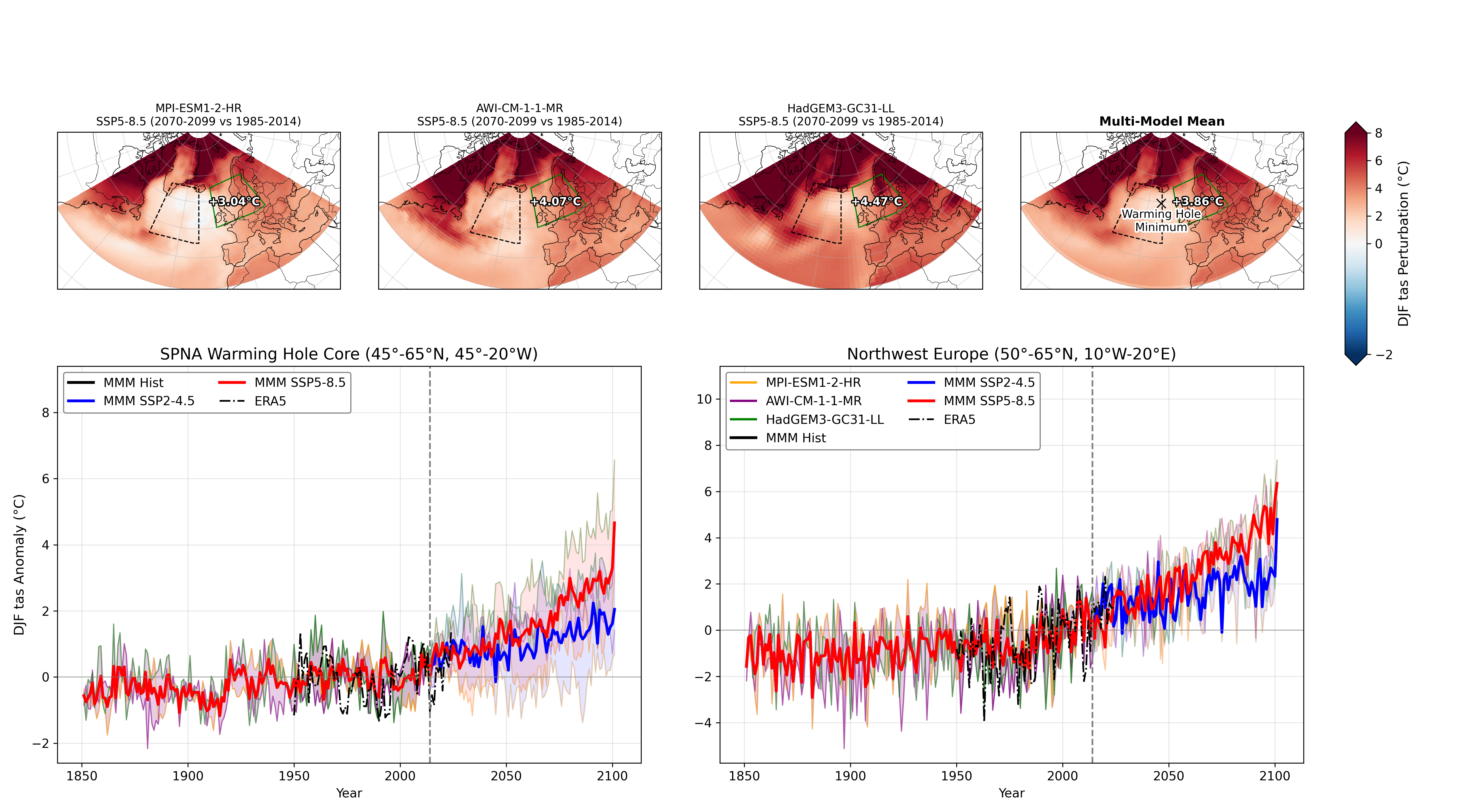}
  \caption{\textbf{Downstream atmospheric response: model-dependent European shielding effect under SSP5-8.5.} Top: DJF surface air temperature perturbation (2070--2099 versus 1985--2014) for the three carry-forward CMIP6 models and the MMM, with the SPNA core (dashed black) and Northwest European (solid green) evaluation boxes overlaid; annotated values give the area-weighted mean warming inside each box. Bottom: DJF temperature anomaly trajectories 1850--2100 for the SPNA warming-hole core (left) and Northwest Europe (right) under historical, SSP2-4.5, and SSP5-8.5 forcing, with ERA5 as the observational reference. The oceanic warming hole is consistent across the ensemble, but the downstream shielding is not: \texttt{HadGEM3-GC31-LL} shields Northwest Europe by $\sim$1\textdegree C against global warming, whereas \texttt{MPI-ESM1-2-HR} shows essentially no shielding and \texttt{AWI-CM-1-1-MR} accelerates European warming.}
  \label{fig:amoc_downstream}
\end{figure*}

Under SSP5-8.5, the multi-model-mean kinematic fingerprint declines monotonically through the 21st century and reaches $-0.66^\circ\mathrm{C}$ as a 2070--2099 average, surpassing the conventional $-0.5^\circ\mathrm{C}$ emergence threshold across the bulk of the late-century window; individual high-fidelity members (e.g.\ \texttt{HadGEM3-GC31-MM}) cross the threshold earlier, near mid-century, while the multi-model-mean does not stay permanently below the threshold within the detection horizon (\cref{fig:amoc_upstream}, panel~b). The downstream atmospheric response, however, refines rather than confirms the canonical shielding narrative (\cref{fig:amoc_downstream}). Quantifying the shielding parameter as the deficit of Northwest European (50\textdegree--65\textdegree N, 10\textdegree W--20\textdegree E) DJF warming relative to the global mean, the agent finds that only \texttt{HadGEM3-GC31-LL} produces a mathematically strong shield ($-0.91^\circ\mathrm{C}$), \texttt{MPI-ESM1-2-HR} collapses to near-neutral ($+0.03^\circ\mathrm{C}$) despite hosting the most pronounced oceanic warming hole, and \texttt{AWI-CM-1-1-MR} in fact warms Northwest Europe \emph{faster} than the global mean ($+0.52^\circ\mathrm{C}$ excess). The ensemble-mean shielding parameter is therefore only $-0.12^\circ\mathrm{C}$, indistinguishable from zero given inter-model spread, while the absolute Northwest European warming reaches $+3.86^\circ\mathrm{C}$ at the MMM. We emphasise that the kinematic fingerprint emergence reported here corresponds to a sustained structural \emph{weakening} of the overturning circulation, not to a bistable shutdown: the synthesis assembled across the agent's chained sessions and the broader CMIP6 consensus place the projected late-century decline at $\sim$35--45\% under SSP5-8.5, and virtually no standard, un-hosed CMIP6 Earth System Model simulates a hysteresis-driven transition below $\sim$5\,Sv before 2100. Recent statistical extrapolations of observational SST fingerprints \citep{Ditlevsen2023} have argued for an earlier tipping window, but these arguments lie outside the CMIP6 standard-ensemble framework and rely on assumptions about early-warning signal stationarity that are themselves contested. The session-level conclusion, that a pronounced SPNA warming hole does not translate into a reliable European cooling buffer because continental thermodynamic heating and shifts in the North Atlantic atmospheric circulation (e.g.\ the projected northeastward migration of the Icelandic Low under SSP5-8.5; \citealp{Huai2025}) can overwhelm the oceanic signal in the carry-forward ensemble, is precisely the kind of cross-domain inference that the integrated literature--data--code workflow is designed to surface. The full 15-model fidelity table, per-scenario shielding values, and raw session telemetry are provided in Section~S1 of the Supplementary Materials.

\FloatBarrier

\subsubsection{ENSO Asymmetry and Extreme Event Projections}\label{sec:uc_enso}

The second use case evaluates the historical fidelity and future evolution of the El Niño--Southern Oscillation (ENSO), the dominant mode of global interannual climate variability. The agent isolated the Niño 3.4 index by computing the area-weighted Sea Surface Temperature (SST) anomaly over the Niño 3.4 region (5\textdegree S--5\textdegree N, 170\textdegree W--120\textdegree W) and detrended the time series using a 30-year rolling climatology, cleanly separating the interannual ENSO signal from the background thermodynamic warming trend. The 10-model ensemble was initially ranked against ERA5 (1950--2014) using a chronological RMSE of the 21-year running standard deviation of the Niño 3.4 index. Following a peer-review intervention (see \cref{sec:discuss_review}), the agent recognised that chronological RMSE unfairly penalises uninitialised models for phase-misaligned internal variability, and replaced it with a \emph{Distribution RMSE} that compares the sorted distributions of the rolling standard deviation against the observational distribution. Under the corrected metric, the high-fidelity ensemble shifts to \texttt{AWI-CM-1-1-MR} (RMSE~$0.018$, skewness~$+0.11$ versus observed $+0.36$), \texttt{FGOALS-g3} (RMSE~$0.021$, skewness~$-0.36$), and \texttt{BCC-CSM2-MR} (RMSE~$0.028$, skewness~$-0.73$). This top-three set carries a notable caveat: the latter two models, despite reproducing observed ENSO amplitude almost exactly, exhibit an \emph{inverted} skewness, a structural symptom of the well-documented Equatorial Cold Tongue Bias, which suppresses the eastward extension of deep convection and damps the nonlinear El Niño/La Niña asymmetry. Amplitude fidelity, in other words, does not guarantee asymmetry fidelity.

\begin{figure*}[!b]
  \centering
  \includegraphics[width=\textwidth]{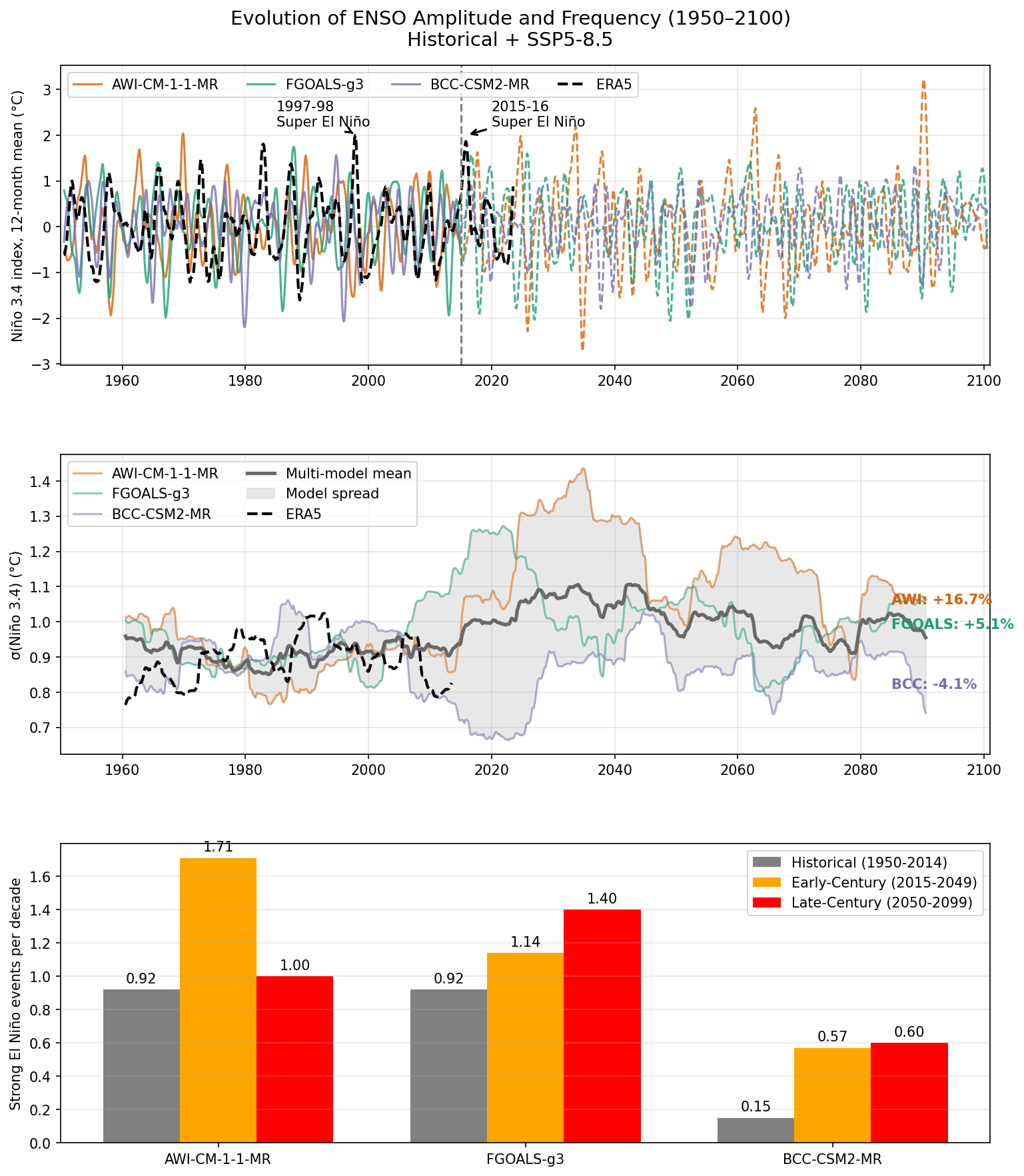}
  \caption{\textbf{Projected evolution of ENSO amplitude and frequency, 1950--2100, under SSP5-8.5 for the three high-fidelity carry-forward models.} Top: 12-month-mean Niño 3.4 index for \texttt{AWI-CM-1-1-MR}, \texttt{FGOALS-g3}, and \texttt{BCC-CSM2-MR}, with ERA5 overlaid; the 1997--98 and 2015--16 super El Niños are annotated for reference. Middle: 21-year running standard deviation $\sigma(\mathrm{Ni\tilde{n}o\,3.4})$ with the multi-model envelope shaded; annotated values are the per-model percentage change in late-century $\sigma$ versus the 1980--2014 baseline ($+16.7\%$, $+5.1\%$, $-4.1\%$). Bottom: decadal frequency of strong El Niño events ($>1.5^\circ\mathrm{C}$ for $\geq$3 consecutive months) for the historical, early-century (2015--2049), and late-century (2050--2099) windows. All three models project a marked rise in the strong-event rate even when the bulk $\sigma$ decreases (the \texttt{BCC-CSM2-MR} pattern), consistent with a fattening of the upper tail of the Niño~3.4 distribution rather than a uniform amplitude scaling.}
  \label{fig:enso_evolution}
\end{figure*}

Projecting the carry-forward ensemble to 2100 under SSP5-8.5 (\cref{fig:enso_evolution}) reveals deep inter-model uncertainty about overall amplitude change, ranging from a $+16.7\%$ intensification in \texttt{AWI-CM-1-1-MR} through $+5.1\%$ in \texttt{FGOALS-g3} to a slight $-4.1\%$ dampening in \texttt{BCC-CSM2-MR}. The frequency of strong El Niño events (Niño~3.4 anomaly $>1.5^\circ\mathrm{C}$ for $\geq$3 consecutive months) tells a more coherent and physically important story. \texttt{AWI-CM-1-1-MR} jumps from 0.92 to 1.71 events per decade by mid-century before relaxing to 1.00 in the late century; \texttt{FGOALS-g3} climbs steadily from 0.92 to 1.40; and, most strikingly, \texttt{BCC-CSM2-MR} quadruples its strong-event rate from 0.15 to 0.60 per decade \emph{despite} its late-century $\sigma$ falling by 4.1\%. This BCC-CSM2-MR pattern is the use case's most diagnostic result: the simultaneous rise in strong-event frequency and decline in bulk $\sigma$ is consistent with a \emph{fattening of the upper tail} of the Niño 3.4 distribution under SSP5-8.5, rather than a uniform amplification of the oscillation, although the present analysis does not include an explicit higher-order tail-shape diagnostic (e.g.\ kurtosis or quantile separation) to discriminate fattening from a shift of the mean state. The full 10-model fidelity ranking, the chronological vs.\ Distribution RMSE diagnostic underlying the peer-review intervention, and the detailed analysis of the cold tongue bias and skewness trap are provided in Section~S2 of the Supplementary Materials.

\FloatBarrier

\subsubsection{Western Boundary Currents and the Internal-Variability Veil}\label{sec:uc_eddy}

\begin{figure*}[!b]
  \centering
  \includegraphics[width=\textwidth,height=0.78\textheight,keepaspectratio]{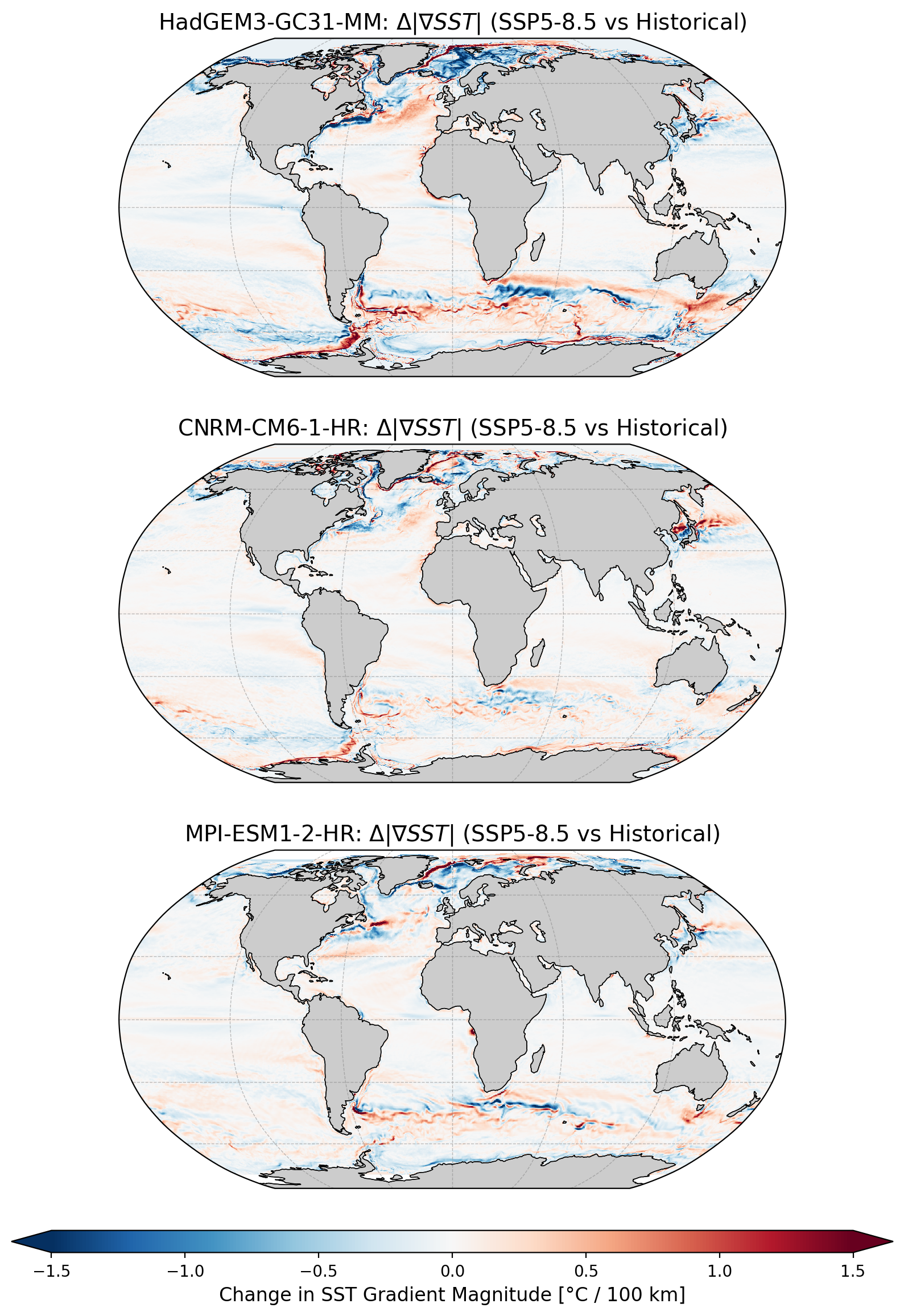}
  \caption{\textbf{Projected change in oceanic frontal sharpness $\Delta|\nabla\mathrm{SST}|$ (late 21st century, SSP5-8.5, versus 1985--2014) for three eddy-permitting CMIP6 configurations.} Robinson-projected global maps for \texttt{HadGEM3-GC31-MM} (top), \texttt{CNRM-CM6-1-HR} (middle), and \texttt{MPI-ESM1-2-HR} (bottom). The diverging colour scale spans $\pm 1.5^\circ\mathrm{C}/100\,\mathrm{km}$; localised dipole structures across the Gulf Stream, Kuroshio Extension, and Antarctic Circumpolar Current reach $\pm 5$--$15^\circ\mathrm{C}/100\,\mathrm{km}$ in the truly eddy-permitting ($\sim$1/4\textdegree) configurations. The dipoles encode the poleward displacement of WBCs (equatorward flank weakening paired with poleward flank sharpening); the coarser \texttt{MPI-ESM1-2-HR} ($\sim$0.4\textdegree) produces a smoother, lower-magnitude response that confirms the resolution dependence of the diagnostic.}
  \label{fig:eddy}
\end{figure*}

The third use case targets the projected response of oceanic mesoscale fronts to greenhouse warming in the major Western Boundary Currents (WBCs), and exposes the limits of single-realisation CMIP6 inference at decadal scales. The agent's literature synthesis identified three competing drivers of mesoscale eddy activity in a warming climate (enhanced stratification damping baroclinic instability, sharpening SST gradients fuelling it, and air-sea coupling exerting substantial mechanical ``eddy killing'' damping on EKE) and then retrieved three eddy-permitting CMIP6 configurations (\texttt{HadGEM3-GC31-MM} and \texttt{CNRM-CM6-1-HR} at $\sim$1/4\textdegree, \texttt{MPI-ESM1-2-HR} at $\sim$0.4\textdegree) to compute the change in SST frontal sharpness, $\Delta|\nabla\mathrm{SST}|$, between the historical (1985--2014) and late-21st-century (2070--2099, SSP5-8.5) climatologies (\cref{fig:eddy}). The eddy-permitting maps reveal localised dipoles of order $\pm 5$--$15^\circ\mathrm{C}/100\,\mathrm{km}$ across the Gulf Stream and Kuroshio Extensions (signatures of poleward WBC displacement), with the strongest extrema in \texttt{HadGEM3-GC31-MM}, while the coarser \texttt{MPI-ESM1-2-HR} produces a visibly smoother response and a broad band of frontal sharpening across the Antarctic Circumpolar Current, consistent with the strengthened Southern Hemisphere westerlies projected by the same ensemble. Native-grid gradient computation, sea-ice-aware masking, and the operator-order choice for the maps, together with the differing regrid-then-gradient choice used in the downstream time-series pipeline, are documented in Section~S3.

A second session evaluated the time-evolving regional-mean frontal sharpness over the Gulf Stream (30\textdegree--45\textdegree N, 80\textdegree--40\textdegree W) and Kuroshio (25\textdegree--45\textdegree N, 140\textdegree--180\textdegree E) extensions 1950--2100 for an ensemble of \texttt{MPI-ESM1-2-HR}, \texttt{EC-Earth3}, and \texttt{CMCC-CM2-SR5}, against an ERA5 1950--2024 reference. The Kuroshio shows a unanimous forced weakening ($-0.07 \pm 0.03\,\mathrm{K}/100\,\mathrm{km}$ for 2071--2100 versus 1985--2014, with all three models negative); the Gulf Stream multi-model mean is $-0.03 \pm 0.08\,\mathrm{K}/100\,\mathrm{km}$ with \emph{no sign agreement} across the ensemble (\texttt{MPI}: $-0.07$; \texttt{EC-Earth3}: $-0.10$; \texttt{CMCC-CM2-SR5}: $+0.08$). Crucially, the recent ERA5 1993--2024 sharpening trend ($+0.020$ to $+0.021\,\mathrm{K}/100\,\mathrm{km}$ per decade in both basins) sits at the edge of, or entirely outside, the simulated free-running model spread over the same window. Because the CMIP6 historical members are uninitialised, their internal decadal variability is out of phase with nature and the multi-model mean smooths it out, leaving only the forced signal; the discrepancy with ERA5 thus admits two readings: either the observed sharpening is a transient internal phase that temporarily masks a long-term forced weakening, or the models structurally fail to capture an emergent forced strengthening mechanism. Either reading carries direct implications for marine ecosystem and storm-track impact projections, and is exactly the type of falsifiable, methodology-aware conclusion that a fully autonomous CMIP6 analysis pipeline should be able to deliver. Full single-member caveats, frontal time-series figures, and per-model trend tables are in Section~S3 of the Supplementary Materials.

\FloatBarrier

\subsubsection{Mediterranean Summer Warming and Sub-Regional Consistency}\label{sec:uc_heatwaves}

The fourth use case targets Mediterranean summer (JJA) temperature projections, a region documented to amplify global warming by approximately a factor of 1.4 and identified by the agent's literature synthesis as a hotspot for compounding soil-moisture/temperature feedbacks, persistent ``double-jet'' Rossby-wave trapping, and Hadley-cell-expansion-driven subsidence. Nine CMIP6 models were evaluated over the Mediterranean domain (30\textdegree--48\textdegree N, 10\textdegree W--40\textdegree E) against ERA5 for the 1985--2014 historical baseline using spatial-pattern RMSE and native area-mean bias. The agent identified \texttt{EC-Earth3} (RMSE $1.35^\circ\mathrm{C}$, bias $+0.08^\circ\mathrm{C}$), \texttt{MPI-ESM1-2-HR}, \texttt{ACCESS-CM2}, \texttt{GFDL-CM4}, and \texttt{CESM2} as the five highest-fidelity members, with \texttt{MIROC6} a profound outlier (RMSE $5.75^\circ\mathrm{C}$, bias $+5.19^\circ\mathrm{C}$). ESGF availability constraints excluded \texttt{HadGEM3-GC31-LL} and forced a sibling-member substitution for \texttt{CNRM-ESM2-1}, both autonomously logged in the session trace (Section~S4).

\begin{figure*}[!b]
  \centering
  \includegraphics[width=\textwidth]{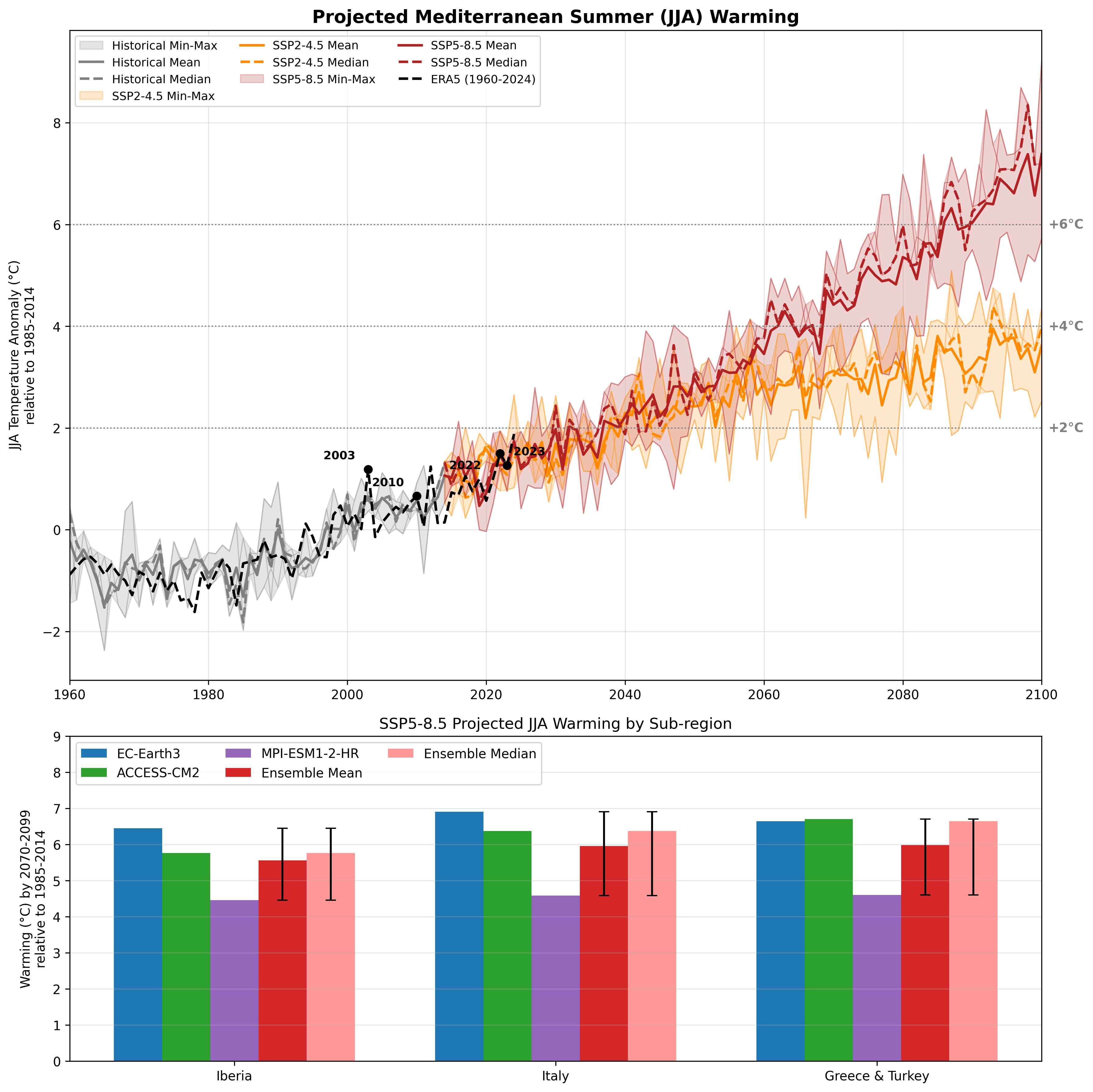}
  \caption{\textbf{Projected Mediterranean summer (JJA) warming, 1960--2100, for the three carry-forward CMIP6 models.} Top: regional JJA temperature anomaly relative to the 1985--2014 baseline, with the ERA5 observational record (1960--2024, dashed black) and explicit annotations of the 2003, 2010, 2022, and 2023 extreme summers; SSP2-4.5 (orange) and SSP5-8.5 (dark red) are shown as ensemble mean (solid), median (dashed), and min--max envelopes. Bottom: sub-regional late-century (2070--2099 versus 1985--2014) warming under SSP5-8.5 for Iberia, Italy, and Greece~\&~Turkey, comparing individual carry-forward models (\texttt{EC-Earth3}, \texttt{ACCESS-CM2}, \texttt{MPI-ESM1-2-HR}) against the ensemble mean and median with error bars. The Mediterranean-wide ensemble-mean warming reaches $+5.67^\circ\mathrm{C}$ under SSP5-8.5; the inter-model spread dwarfs the inter-region differences.}
  \label{fig:heatwaves}
\end{figure*}

A subsequent session applied a rank-product rule combining absolute bias and spatial RMSE to select \{\texttt{EC-Earth3}, \texttt{ACCESS-CM2}, \texttt{MPI-ESM1-2-HR}\} as the carry-forward ensemble for SSP-scenario projections (\cref{fig:heatwaves}). The Mediterranean-wide ensemble-mean JJA warming for 2070--2099 versus 1985--2014 reaches $+5.67^\circ\mathrm{C}$ under SSP5-8.5, with sub-regional means of $+5.55^\circ\mathrm{C}$ (Iberia), $+5.96^\circ\mathrm{C}$ (Italy), and $+5.98^\circ\mathrm{C}$ (Greece~\&~Turkey). Following an explicit user-level instruction, the agent reported a crucial consistency caveat that conventional ensemble-mean reporting suppresses: the inter-member spread across the three carry-forward models exceeds $2^\circ\mathrm{C}$ at the sub-regional scale (e.g.\ \texttt{MPI-ESM1-2-HR} projects $\sim$4.6\textdegree C while \texttt{ACCESS-CM2} projects $\sim$6.7\textdegree C over Greece~\&~Turkey), whereas the apparent geographic gradient between sub-regions is $\sim 0.4^\circ\mathrm{C}$. The sub-regional ranking is therefore not stable against single-realisation internal decadal variability, and the headline ensemble-mean number is the only statistically defensible quantity at this ensemble size. In the ERA5 observational record, the 2003, 2022, and 2023 Mediterranean summers each exceed $+1^\circ\mathrm{C}$ above the 1985--2014 baseline (namely $+1.18^\circ\mathrm{C}$, $+1.50^\circ\mathrm{C}$, and $+1.26^\circ\mathrm{C}$), while the 2010 summer reaches only $+0.66^\circ\mathrm{C}$ over the Mediterranean basin as a whole; all four sit within the bulk of the historical model envelope. Full evaluation tables, the spatial-pattern figure, the rank-product computation, and the \texttt{xesmf}-package-unavailable fallback to Xarray bilinear interpolation are in Section~S4 of the Supplementary Materials.

\FloatBarrier

\subsubsection{North Atlantic Oscillation: Spatial Migration Versus Internal Variability}\label{sec:uc_nao}

\begin{figure*}[!b]
  \centering
  \includegraphics[width=\textwidth]{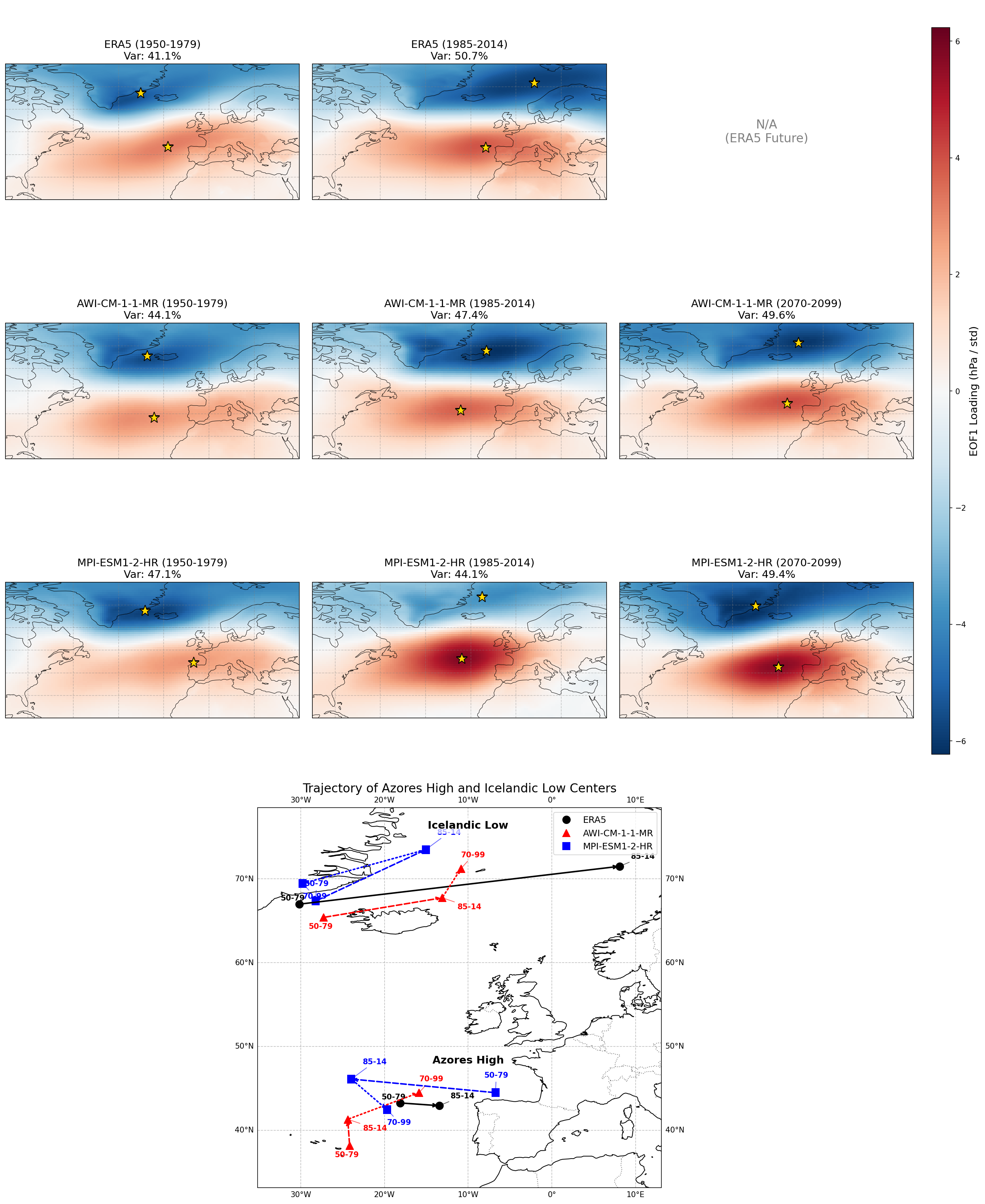}
  \caption{\textbf{Structural evolution of the North Atlantic Oscillation across three 30-year windows.} Top: EOF1 spatial loading of DJF sea-level pressure (hPa per standardised principal-component unit) for ERA5 (top row), \texttt{AWI-CM-1-1-MR} (middle row), and \texttt{MPI-ESM1-2-HR} (bottom row), across 1950--1979, 1985--2014, and 2070--2099 under SSP5-8.5 (top-right ERA5 cell is unavailable). Gold stars mark the area-weighted centroids of the Icelandic Low and Azores High in each epoch; the explained variance is annotated. Bottom: trajectory map summarising the centroid migration across the three epochs for ERA5 (black circles), \texttt{AWI-CM-1-1-MR} (red triangles), and \texttt{MPI-ESM1-2-HR} (blue squares). Both models qualitatively reproduce the observed eastward historical migration of the Icelandic Low but disagree on its projected late-21st-century trajectory, exemplifying the dominance of internal decadal variability over forced response at the single-realisation 30-year sample size.}
  \label{fig:nao}
\end{figure*}

The fifth use case demonstrates an explicit failure mode of single-realisation CMIP6 inference for atmospheric modes of variability \citep{McKenna2021,Eade2024}. The agent computed the first Empirical Orthogonal Function (EOF1) of December--February (DJF) sea-level pressure over the North Atlantic for three 30-year windows (1950--1979, 1985--2014, and 2070--2099 under SSP5-8.5) and extracted the area-weighted centroids of the Icelandic Low and Azores High to trace the spatial migration of the NAO's action centres (\cref{fig:nao}). ERA5 shows a striking $\sim 38^\circ$ eastward displacement of the Icelandic Low between the two historical windows (from 30.1\textdegree W off Greenland to 8.1\textdegree E in the Norwegian Sea), with the Azores High shifting modestly eastward (from 18.1\textdegree W to 13.5\textdegree W). Both \texttt{AWI-CM-1-1-MR} and \texttt{MPI-ESM1-2-HR} qualitatively reproduce the observed eastward migration of the Icelandic Low between the two historical epochs, but the simulated magnitudes are roughly a third of the observed displacement ($\sim 14^\circ$ and $\sim 13^\circ$ respectively), and the response of the Azores High disagrees between the two models, with \texttt{AWI} shifting slightly northward and \texttt{MPI} retreating westward.

Under SSP5-8.5 the two single-realisation projections diverge in sign for the Icelandic Low displacement: \texttt{AWI-CM-1-1-MR} projects a continued northeastward displacement of both centres (Icelandic Low to 10.8\textdegree W, 71.2\textdegree N; Azores High to 15.9\textdegree W, 44.5\textdegree N), broadly consistent with the more-positive-and-less-variable NAO response identified across CMIP6 at high CO$_2$ forcing \citep{Mitevski2025}, whereas \texttt{MPI-ESM1-2-HR} projects the Icelandic Low retreating westward back to 29.8\textdegree W, near its mid-twentieth-century position, with a concurrent southward shift. The agent itself framed the central methodological conclusion: a 30-year, single-realisation EOF centroid cannot confidently separate forced response from low-frequency internal variability of the NAO, which itself modulates the North Atlantic tripole SST on interdecadal scales \citep{Song2024}, and the observed $\sim 38^\circ$ ERA5 displacement is most likely dominated by the latter rather than by a forced poleward expansion of the storm track. Used naively, this analysis would license a confidently wrong claim about NAO migration under climate change. The use case is consequently as much an illustration of the system's epistemic honesty (explicitly bounding its own inference and calling for larger initial-condition ensembles, i.e.\ Single-Model Initial-Condition Large Ensembles or SMILEs, for a defensible attribution) as it is of its analytical capacity. Full action-centre trajectory tables and EOF-variance-explained statistics are in Section~S5 of the Supplementary Materials.

\FloatBarrier

\subsubsection{Three Precipitation Regimes Under SSP5-8.5}\label{sec:uc_precip}

The sixth use case targets regional precipitation under high-emission forcing and demonstrates the agent's ability to distinguish prognostically reliable from structurally uncertain CMIP6 projections in a single, integrated session. Ten CMIP6 models were evaluated against ERA5 (1985--2014) on two complementary metrics: terrestrial spatial RMSE of the annual-mean precipitation field, and an ITCZ latitudinal bias diagnosed as the precipitation-weighted mean latitude over the tropical Pacific (10\textdegree S--10\textdegree N, 160\textdegree E--90\textdegree W), designed to isolate the well-documented double-ITCZ failure mode. Against an ERA5 baseline of $2.91\,\mathrm{mm\,day^{-1}}$ global mean and a precipitation-weighted mean ITCZ latitude of $2.30^\circ$N, the agent selected \texttt{NorESM2-MM} (RMSE $1.216\,\mathrm{mm\,day^{-1}}$, ITCZ bias $-0.08^\circ$), \texttt{GFDL-CM4} ($1.238$, $-0.54^\circ$), and \texttt{EC-Earth3} ($1.240$, $-0.97^\circ$) as the carry-forward ensemble. Land-fraction handling, grid harmonisation across cubed-sphere and Gaussian native geometries, and the full 10-model evaluation table are in Section~S6.

\begin{figure*}[!b]
  \centering
  \includegraphics[width=\textwidth]{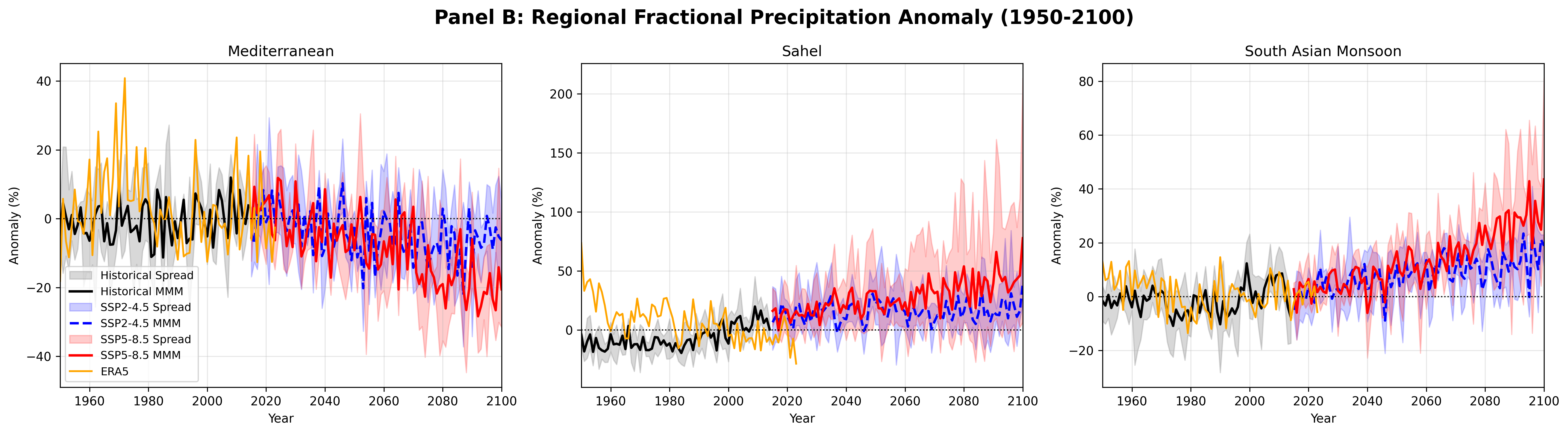}
  \caption{\textbf{Three regional precipitation regimes under SSP5-8.5, 1950--2100.} Time series of the fractional precipitation anomaly relative to the 1985--2014 baseline for the three carry-forward models' multi-model mean over the Mediterranean (left), Sahel transition zone (centre), and South Asian Monsoon (right). The historical MMM is shown in black, SSP2-4.5 MMM as a dashed blue line, SSP5-8.5 MMM as a solid red line, and the ERA5 baseline as an orange line over the observed period; shaded envelopes give the inter-model min--max spread. The Mediterranean exhibits structurally consistent drying ($-18.5\%$ under SSP5-8.5, with a tight envelope), the South Asian Monsoon exhibits consistent wetting ($+24.8\%$, with all members positive), and the Sahel envelope explicitly crosses zero, encoding structural inter-model disagreement.}
  \label{fig:precip}
\end{figure*}

Late-century projections (2070--2099 versus 1985--2014) for the three carry-forward models partition into three qualitatively distinct prognostic regimes (\cref{fig:precip}). The \textbf{Mediterranean} basin (30\textdegree--45\textdegree N, 10\textdegree W--35\textdegree E) exhibits structurally systematic drying of $-18.49\%$ ($-0.22\,\mathrm{mm\,day^{-1}}$) under SSP5-8.5 with an exceptionally tight inter-model spread of $-0.27$ to $-0.19\,\mathrm{mm\,day^{-1}}$, consistent with Hadley-cell-expansion-driven subsidence and persistent anticyclonic forcing; the same drying signal is already $-7.04\%$ under SSP2-4.5. The \textbf{South Asian Monsoon} domain (5\textdegree--30\textdegree N, 60\textdegree--100\textdegree E) shows the converse: systematic wetting of $+24.79\%$ ($+0.70\,\mathrm{mm\,day^{-1}}$) under SSP5-8.5 with all three models positive (spread $+0.38$ to $+1.04\,\mathrm{mm\,day^{-1}}$), driven by Clausius--Clapeyron moisture enhancement and an extended monsoon withdrawal phase. The \textbf{Sahel transition zone} (10\textdegree--20\textdegree N, 20\textdegree W--40\textdegree E) provides the prognostic-ambiguity counter-example: the multi-model-mean fractional shift is $+39.03\%$, but the absolute inter-model spread of $-0.02$ to $+0.56\,\mathrm{mm\,day^{-1}}$ explicitly crosses the zero line, indicating that the CMIP6 ensemble structurally disagrees on whether the region will face wetting or drying under high-emission forcing. Reporting the bare ensemble mean of $+39\%$ without this spread would be substantively misleading. Full evaluation, projection tables, and the precipitation climatology evaluation figure are in Section~S6 of the Supplementary Materials.

\FloatBarrier

\subsubsection{Constrained-Ensemble GMST Projections and Paris-Threshold Crossings}\label{sec:uc_warming}

The seventh and final use case targets the central diagnostic of CMIP6 evaluation, global mean surface temperature (GMST), and exercises the system's ability to operationalise the well-documented ``hot model'' problem of the CMIP6 generation. The agent's literature synthesis identified that high-Equilibrium-Climate-Sensitivity ($\mathrm{ECS}>4.5\,\mathrm{K}$) configurations, driven primarily by overly positive extratropical cloud feedbacks, systematically overestimate the post-1990 observed warming and inflate the upper tail of 21st-century projections; the agent therefore restricted the evaluation a priori to ten low-to-medium-ECS members ($\mathrm{ECS}\leq 4.5\,\mathrm{K}$). Two RMSE diagnostics were computed against ERA5 anomalies relative to 1981--2010: a chronological RMSE over 1980--2014, which is dominated by phase mismatches in unforced internal variability for free-running members, and a decadal RMSE over 1980--2010 after a 10-year centred rolling smoother, which isolates the forced response. The decadal smoother universally reduced the RMSE relative to the chronological value, confirming that for forced-response evaluation the smoothed metric is the physically meaningful choice. The agent's reported and projected ensemble was \texttt{BCC-CSM2-MR} (Decadal RMSE $0.051\,\mathrm{K}$), \texttt{GFDL-ESM4} ($0.049$), \texttt{MIROC6} ($0.055$), \texttt{MPI-ESM1-2-HR} ($0.066$), and \texttt{MRI-ESM2-0} ($0.061$). A strict decadal-RMSE ranking of the full 10-model candidate set places \texttt{MPI-ESM1-2-HR} seventh rather than fourth and would substitute \texttt{NorESM2-LM} ($0.063\,\mathrm{K}$) in the fifth slot; the projections reported below are nevertheless computed on the agent's selected five-member ensemble, with the consequences of this ranking artefact for the late-century crossing years documented in the Telemetry Provenance section and in Section~S7 of the Supplementary Materials.

\begin{figure*}[!b]
  \centering
  \includegraphics[width=\textwidth]{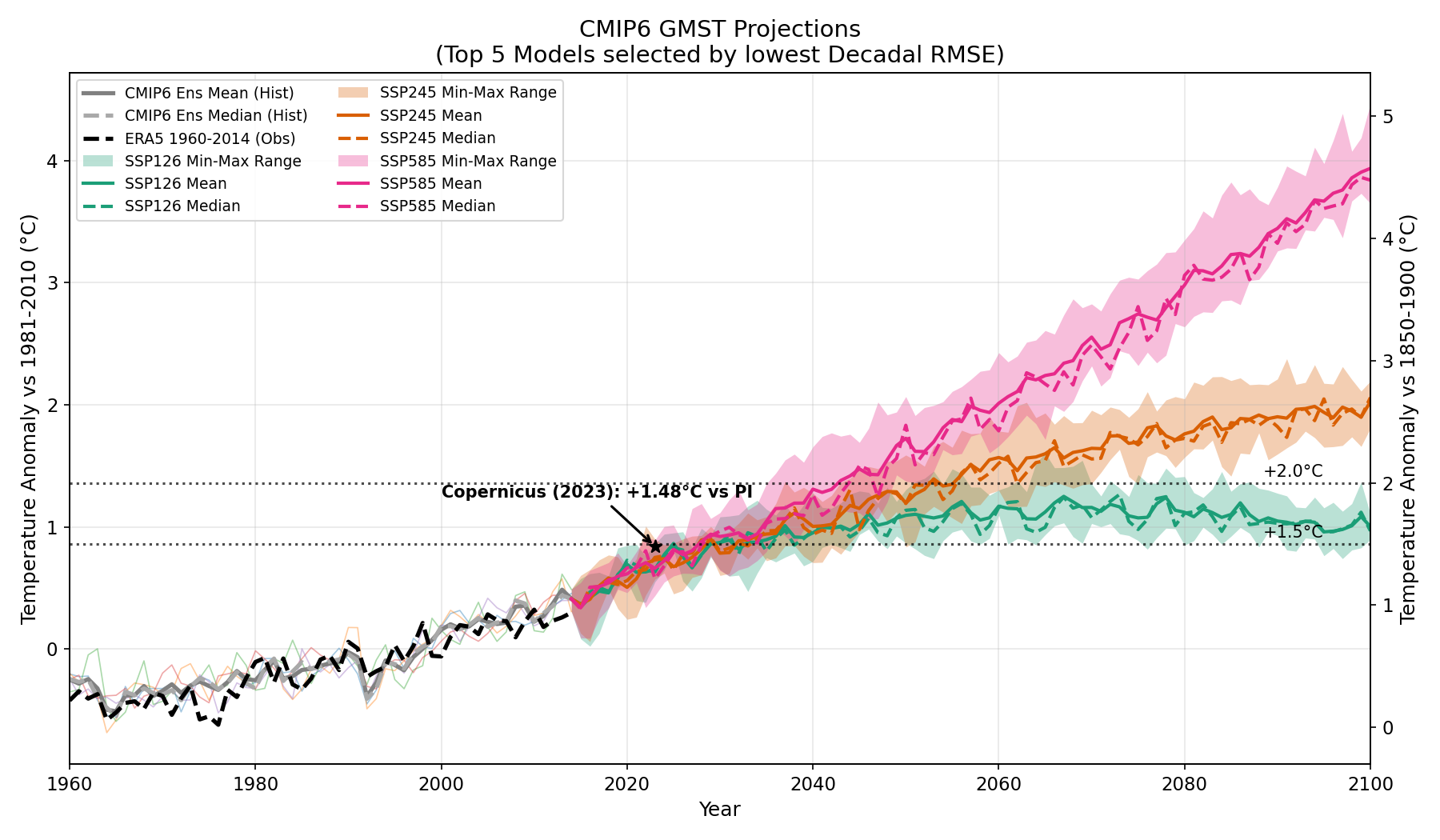}
  \caption{\textbf{Constrained-ensemble GMST projections, 1960--2100, under three SSP scenarios.} The five CMIP6 models selected by the agent in the historical evaluation (\texttt{BCC-CSM2-MR}, \texttt{GFDL-ESM4}, \texttt{MIROC6}, \texttt{MPI-ESM1-2-HR}, \texttt{MRI-ESM2-0}) are projected under SSP1-2.6 (green), SSP2-4.5 (orange), and SSP5-8.5 (magenta), with ensemble mean (solid), median (dashed), and min--max envelopes (shaded). The left axis is the anomaly relative to 1981--2010; the right axis is offset by $+0.639^\circ\mathrm{C}$ (computed from NASA GISTEMP v4, consistent with AR6 WG1 Box 2.3) to express anomalies relative to the 1850--1900 pre-industrial baseline. Horizontal dotted lines mark the $+1.5^\circ\mathrm{C}$ and $+2.0^\circ\mathrm{C}$ Paris-Agreement thresholds; the Copernicus 2023 annual anomaly of $+1.48^\circ\mathrm{C}$ versus pre-industrial is annotated.}
  \label{fig:gmst}
\end{figure*}

This constrained five-member ensemble was projected under SSP1-2.6, SSP2-4.5, and SSP5-8.5 to 2100 (\cref{fig:gmst}). To translate the ensemble onto the Paris-Agreement frame the agent computed an explicit pre-industrial offset of $+0.639^\circ\mathrm{C}$ between the 1981--2010 reference and the 1850--1900 baseline approximated by the 1880--1900 GISTEMP~v4 record (since the GISTEMP dataset begins in 1880); the value is consistent with the AR6 WG1 Box 2.3 estimate. Crossing years for the $+1.5^\circ\mathrm{C}$ and $+2.0^\circ\mathrm{C}$ thresholds were defined per AR6 Cross-Chapter Box 11.1 methodology using a 20-year centred running mean, with the resulting boundary truncation limiting detection to year 2090. The constrained ensemble mean crosses $+1.5^\circ\mathrm{C}$ in 2029 under SSP5-8.5, 2032 under SSP2-4.5, and 2033 under SSP1-2.6; $+2.0^\circ\mathrm{C}$ is crossed in 2044 under SSP5-8.5 and 2054 under SSP2-4.5, while SSP1-2.6 does not cross $+2.0^\circ\mathrm{C}$ before the 2090 detection horizon (only \texttt{MRI-ESM2-0} briefly does in 2058). The Copernicus 2023 annual mean of $+1.48^\circ\mathrm{C}$ relative to 1850--1900 sits within the bulk of the constrained ensemble at the same date. Full decadal/chronological RMSE tables, ECS values, member-level crossing distributions, and the historical evaluation figure are in Section~S7 of the Supplementary Materials.

\FloatBarrier

\section{Discussion and Conclusions}\label{sec:discussion}

\subsection{Autonomous Adversarial Peer Review}\label{sec:discuss_review}

CMIP-Forge implements an \textbf{Autonomous Adversarial Peer Review Protocol} to address the well-documented susceptibility of LLM-generated scientific code to silent methodological failures, such as unit conversion errors, calendar mismatches, and incorrect spatial aggregations. By subjecting the worker agent's analytical pipelines to independent critiques from multiple foundation models, the system establishes a self-correcting feedback loop.

However, systematic evaluation across extended use cases (documented in the Supplementary Materials) has revealed that the current peer review architecture carries significant epistemic risks. Reviewer agents, while capable of generating domain-relevant critiques, can also hallucinate plausible-sounding but factually incorrect physical constraints. When two independent reviewers converge on the same false critique, the worker agent exhibits \textbf{LLM sycophancy}: it defers to the authoritative tone of the reviewers and modifies correct code, introducing errors that would not have occurred without the review mechanism. This ``socially-induced regression'' is qualitatively more dangerous than standard code generation errors, precisely because it originates from a mechanism designed to \emph{improve} reliability.

A complementary failure mode is the \textbf{governance-level bypass} of the review loop itself. The session traces released alongside this article document two distinct variants. First, in the AMOC, NAO, Eddy/WBC, and GMST use cases, the worker agent acknowledged at least one reviewer \texttt{REVISE} verdict requiring a code change but did not formally re-submit the corrected code for a second review pass; the session was closed by the worker rather than by an \texttt{ACCEPT} verdict from the reviewer panel. Second, in the precipitation use case, the worker submitted to both reviewers a placeholder code stub (library imports followed by the literal comment ``\texttt{\# \ldots\ code generated the Panel A and Panel B successfully \ldots}'') in place of the actual data-processing pipeline; both reviewers correctly issued \texttt{REJECT} verdicts on the grounds that the submitted code was not an executable analysis, and the worker proceeded directly to the final response without re-submitting the real pipeline. The actual numerical analysis in that case had been performed correctly in earlier steps within the same session, so the published figures and tables derive from real computation rather than from the stub; the failure is governance-level, not numerical, but it removes the adversarial review safety net for the affected step. Both bypass variants motivate the deterministic-handoff mitigations listed below.

These findings carry fundamental implications for multi-agent system design. Current LLM architectures lack effective mechanisms for epistemic self-confidence: when a worker agent produces correct output that contradicts plausible-sounding external critique, it cannot introspect on the source of its own correctness. Potential mitigations include: (i)~requiring reviewer agents to provide verifiable evidence (e.g., exact API documentation URLs or computed sanity-check values) alongside critiques; (ii)~implementing mandatory ``numerical sanity checks'' that flag order-of-magnitude deviations from physical baselines after any pipeline modification; and (iii)~introducing a dedicated ``arbiter'' agent that mediates worker-reviewer disagreements by executing independent micro-experiments rather than adjudicating on rhetorical authority alone.

\subsection{Integrated ERA5 Reanalysis}\label{sec:era5}

A distinguishing feature of CMIP-Forge is the direct integration of Copernicus ERA5 reanalysis data \citep{Hersbach2020} through the Climate Data Store (CDS) API.
Rather than relying on external agents for observational validation, the system's built-in ERA5 retrieval tool enables the worker agent to autonomously download monthly-averaged reanalysis fields at 0.25\textdegree{} resolution, compute observational climatologies, and evaluate CMIP6 model output against satellite-era constraints within a single analytical session. To ensure reliable caching and prevent collisions between different seasonal queries, the retrieval module incorporates a cryptographic hash of the requested months directly into the filename.
This tight coupling between literature knowledge and observational data enables workflows that would otherwise require manual coordination: for instance, the agent can query its RAG corpus for documented temperature biases in a specific model's historical simulation, then immediately retrieve the corresponding ERA5 fields for quantitative comparison, and apply the appropriate bias correction procedure identified in the literature.

\subsection{Concluding Remarks}\label{sec:conclusion}

We have presented CMIP-Forge, a hybrid RAG and agentic analysis platform that couples a curated corpus of 6{,}581 CMIP6-related publications with a tool-augmented worker and an adversarial reviewer loop, delivering end-to-end climate science workflows under explicit methodological supervision.
The system demonstrates that grounding LLM agents in domain-specific scientific literature, while simultaneously providing access to cloud-hosted climate data and computational tools, enables a qualitatively new mode of scientific inquiry.
Across seven use cases spanning ocean dynamics, atmospheric teleconnections, regional extremes, and global warming projections, a single natural-language prompt was sufficient to trigger substantial end-to-end pipelines that retrieved peer-reviewed literature, located and downloaded the relevant CMIP6 model output, executed quantitative analysis in a sandboxed Python environment, and produced publication-quality figures, while several sessions also surfaced concrete reviewer-loop failure modes that we catalogue without redaction in the Telemetry Provenance section of the Supplementary Materials.

The autonomous adversarial peer review protocol provides a critical quality assurance layer, though our systematic evaluation reveals that current LLM-based review architectures carry significant epistemic risks (including sycophantic capitulation and hallucinated physical constraints) that must be addressed before such systems can be deployed for unsupervised scientific production.

Several limitations bound the system's current capabilities and define the immediate research agenda. First, retrieval performance is fundamentally constrained by the quality and completeness of the underlying corpus: although 6{,}581 papers represent substantial coverage, the CMIP6 literature continues to grow and the current corpus is a snapshot as of early 2026. Second, the chunk-level retrieval granularity, while effective for locating specific claims and methodological details, can miss document-level arguments that span multiple sections; hierarchical retrieval strategies combining chunk-level precision with paper-level summary representations are an obvious next step. Third, because the system relies on stochastic LLM inference, the natural-language outputs are inherently non-deterministic: while the deterministic RAG retrieval and the Python sandbox guarantee that the evidence base and any generated code produce identical output across reruns, the upstream reasoning layer introduces variability in how evidence is synthesised. Fourth, the use cases reported here surface concrete reviewer-loop failure modes (sycophantic regression, hallucinated physical objections, occasional code-stub submission) that motivate the mitigations outlined in Section~4.1 but do not yet implement them; closing this loop is the most important deployment-readiness milestone.

As the climate science community scales toward CMIP7, the traditional manual paradigm of data discovery, literature review, and analysis scripting is becoming cognitively intractable. Hybrid systems like CMIP-Forge bridge this gap, merging unstructured peer-reviewed intelligence with live ESGF data pipelines, offering a pathway from passive knowledge archives to active, literature-informed scientific reasoning engines.

\section*{Data Availability}

The CMIP6 model output used in the use case demonstrations is openly available through the Pangeo Cloud data catalog (\url{https://pangeo-data.github.io/pangeo-cmip6-cloud/}).
The literature corpus is derived from open-access publications.

\section*{Code Availability}

The CMIP-Forge source code is openly available at \url{https://github.com/CliDyn/cmip6_gpt}.

\section*{Acknowledgements}

This work was supported by the Helmholtz Association and the Federal Ministry of Research, Technology and Space (BMFTR) through the DataHub Initiative of the Research Field Earth and Environment; the European Union's Destination Earth Initiative and relates to tasks entrusted by the European Union to the European Centre for Medium-Range Weather Forecasts implementing part of this Initiative with funding by the European Union; and Project S1: Diagnosis and Metrics in Climate Models of the Collaborative Research Centre TRR~181 ``Energy Transfer in Atmosphere and Ocean,'' funded by the Deutsche Forschungsgemeinschaft (DFG, German Research Foundation, project no.~274762653).
This work was also supported by the TerraDT (Digital Twin of Earth system for Cryosphere, Land surface and related interactions) project, which has received funding from the European Union's Horizon Europe research and innovation programme under Grant Agreement no.~101187992.
Views and opinions expressed are those of the authors only and do not necessarily reflect those of the European Union or the European Climate Infrastructure and Environment Executive Agency (CINEA). Neither the European Union nor the granting authority can be held responsible for them.

\bibliographystyle{apalike}

\end{document}